%% file: ms.tex
\documentclass[iop]{emulateapj}
\usepackage{float}
\usepackage{CJK}
\begin{document}

\title{Constraining the Oblateness of {\em Kepler} planets}
\begin{CJK*}{UTF8}{gkai}

\author{Wei Zhu (祝伟) \altaffilmark{1}, Chelsea X. Huang (黄煦) \altaffilmark{2}, George Zhou \altaffilmark{3}and D.N.C. Lin \altaffilmark{4,5}}
\affil{$^1$ Department of Astronomy, The Ohio State University, 140 W. 18th Ave., Columbus, OH 43210, USA, weizhu@astronomy.ohio-state.edu}
\affil{$^2$ Department of Astrophysical Sciences, Princeton University}
\affil{$^3$ Research School of Astronomy and Astrophysics, Australian
National University, Cotter Rd, Weston Creek, ACT 2611, Australia}
\affil{$^4$ UCO/Lick Observatory, University of California, USA }
\affil{$^5$ Institute for Advanced Studies, Tsinghua University, Beijing, China. }

\begin{abstract}
We use {\em Kepler} short cadence light curves to constrain the oblateness of planet candidates in the {\em Kepler} sample. The transits of rapidly rotating planets that are deformed in shape will lead to distortions in the ingress and egress of their light curves. We report the first tentative detection of an oblate planet outside of the solar system, measuring an oblateness of $0.22_{-0.11}^{+0.11}$ for the $18\,M_J$ mass brown dwarf Kepler 39b (KOI-423.01). We also provide constraints on the oblateness of the planets (candidates) HAT-P-7b, KOI-686.01, and KOI-197.01 to be $<0.067$, $<0.251$, and $<0.186$, respectively. Using the $Q^\prime$-values from Jupiter and Saturn, we expect tidal synchronization for the spins of HAT-P-7b, KOI-686.01 and KOI-197.01, and for their rotational oblateness signatures to be undetectable in the current data. The potentially large oblateness of KOI-423.01 (Kepler 39b) suggests that the $Q^\prime$-value of the brown dwarf needs to be two orders of magnitude larger than that of the solar system gas giants to avoid being tidally spun-down.
\end{abstract}

\section{Introduction}

The Solar System planets are oblate in shape due to their rapid rotations. The equatorial radius is larger than the polar radius by 7\,\% for Jupiter, and by 10\,\% for Saturn. The bulk rotational angular momentum of a planet is retained from gas accretion in the formation process \citep[e.g.][]{Lissauer:1995}. For a planet in hydrostatic equilibrium, the level of deformation is related to the rotation rate and moment of inertia of the planet, which in turn are influenced by its density profile \citep{Hubbard:1989,Murray:1999,Carter:2010a}. Measuring the oblateness of exoplanets will enable us to better understand their interior structure, as well as their formation and evolutionary history.

Previous works \citep{Hui:2002,Seager:2002,Barnes:2003,Carter:2010a,Carter:2010b} explored the predicted transit light curves of oblate planets. The transit light curve of an oblate planet a) is asymmetric about the ingress and egress regions, and b) will exhibit small differences over ingress and egress to the transit of a spherical planet. However, it is difficult to photometrically measure the oblateness of an extrasolar planet due to its small amplitude signature. For an optimal case that a transiting Jupiter with planet-to-star radius ratio of 0.15 and a Saturn-like oblateness of 0.1, the maximum deviation between an oblate and a spherical planet transit is 400 parts per million (ppm) over the ingress and egress light curve regions \citep{Carter:2010a}.

The only implementation of oblate transit models to date was carried out by \citet{Carter:2010a}, with seven transits of the hot Jupiter HD~189733b observed by the {\em Spitzer Space Telescope}. They were able to rule out a Saturn-like oblateness for the planet. HD~189733b and other hot-Jupiters, with orbital semi-major axes $<0.2$\,AU, are expected to have spun-down due to tidal dissipation to be tidally locked. The photometric signatures due to oblateness from these slowly rotating gas giants are too small to be measurable by any current facility.

The unprecedented photometric precision of the \emph{Kepler} mission has enabled the detections of many subtle photometric effects, such as phase curves of reflected light from planets \citep{Borucki:2009}, Doppler beaming \citep{Mazeh:2012}, and gravity induced asymmetry due to system spin-orbit misalignments \citep{Barnes:2011,ZhouHuang:2013}. The four-year baseline of \emph{Kepler} also means that observations of planetary transits are no longer restricted to short-period systems. The possibility of observing the transits of gas giants that are not tidally affected and thus still retain their primordial spin rate motivates a new search for oblateness signals in transit light curves. This work is the first survey to use \emph{Kepler} short cadence photometry to measure the oblateness of transiting planets.

The structure of this paper is as follows: \S2 describes the light curve modeling of a transiting oblate planet, and discusses the detectability of such a signal. \S3 describes the target selection and \emph{Kepler} light curve reduction procedures. We apply the oblate planet transit model to \emph{Kepler} light curves in \S4, first as a signal injection and recovery exercise to demonstrate the detectability of the signal, then to model four selected \emph{Kepler} planet candidates. In \S5 we discuss the implications of our results.

\section{Light Curve Model}
\subsection{The Transit of an Oblate Planet}

The shape of an oblate planet can be quantified by the flattening, or oblateness, parameter $f$,
\begin{equation} \label{eq:flattening}
f = \frac{R_{\rm eq}-R_{\rm pol}}{R_{\rm eq}} \, ,
\end{equation}
where $R_{\rm eq}$ and $R_{\rm pol}$ are the equatorial and polar radii, respectively \citep{Murray:1999}. We also define the effective mean radius to be 
\[ R_{\rm p} \equiv \sqrt{R_{\rm eq} R_{\rm pol}}, \] 
which is the radius of a spherical planet of the same cross-sectional area. The planet's spin obliquity, $\theta$, is defined as the angle between the polar axis of the planet and the orbital angular momentum vector \citep{Carter:2010a}.

The flattening $f$ can be related to the rotation period $P_{\rm rot}$ of a planet with mass $M_p$ by 
\begin{equation} \label{eq:rotation}
P_{\rm rot} = 2\pi \sqrt{\frac{R_{\rm eq}^3}{GM_p (2f-3J_2)}} \ ,
\end{equation}
where $J_2$ is the quadrupole moment \citep{Carter:2010a}.

\begin{figure*}
\centering
\plotone{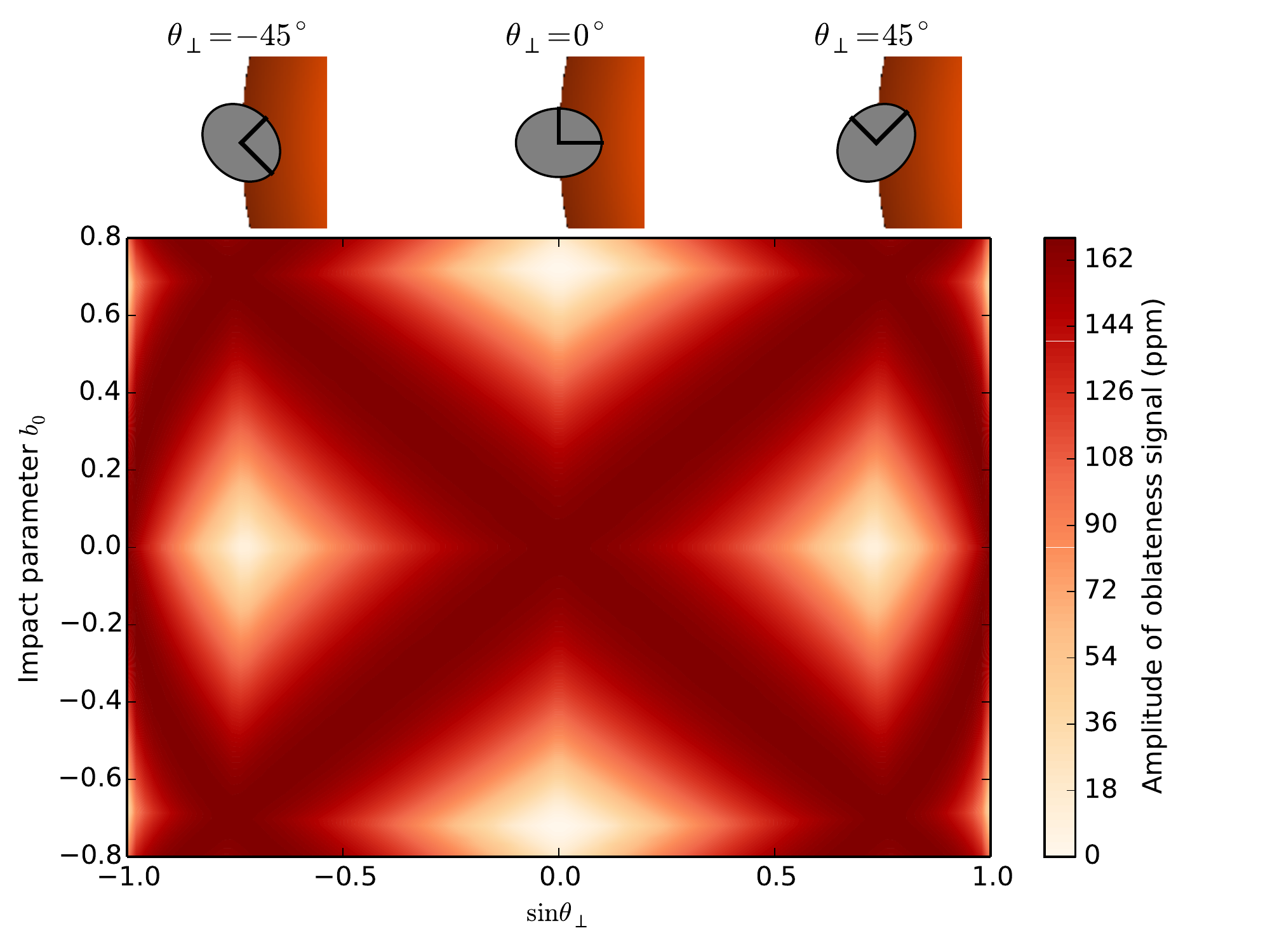}
\caption{Upper panel: three schematic system configurations to demonstrate the definition of $\theta_\perp$; the shape and the relative size of the planet are not scaled for a realistic planet. Lower panel: The amplitude of oblateness signal as a function of the impact parameter $b_0$ and the projected obliquity $\theta_\perp$; a Jupiter-size ($R_{\rm p}/R_\star=0.1$) planet with a Saturn-like oblateness ($f_\perp=0.1$) is assumed.
\label{fig:b0-sina}}
\end{figure*}

However, what we can measure from transit is not the true flattening and obliquity, but their projected components on the plane of the sky, $f_{\perp}$ and $\theta_{\perp}$, meaning that the constrained oblateness from transit is only a lower limit on the true oblateness. The relations between $f$, $\theta$ and $f_{\perp}$, $\theta_{\perp}$ can be found in \citet{Carter:2010a}. The definition for the sign of $\theta_\perp$ is shown in the upper panel of Figure~\ref{fig:b0-sina}.

\begin{figure*}
\plotone{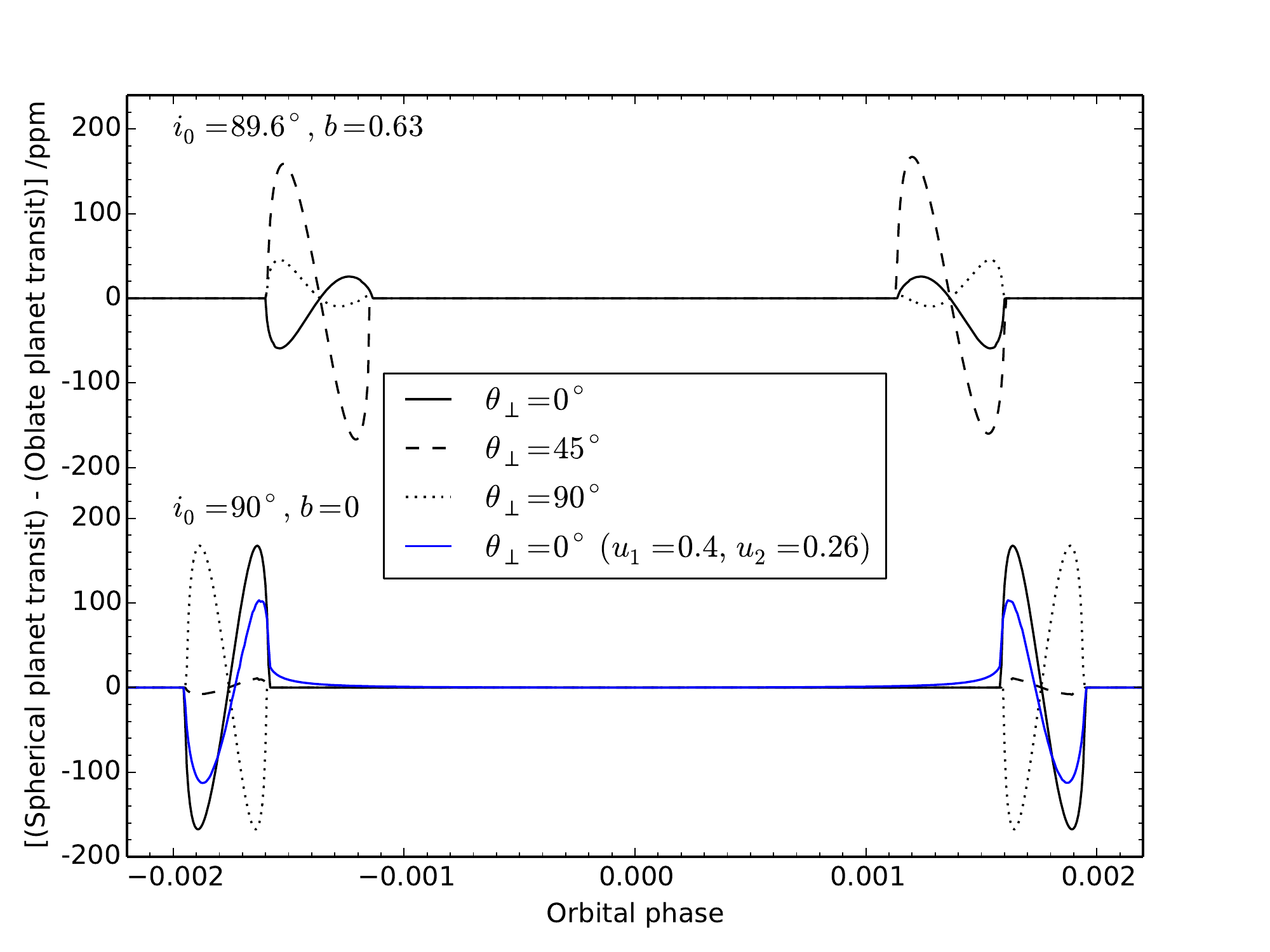}
\caption{Example signals induced by an oblate planet with respect to a spherical planet of the same cross-sectional area, for $\theta_\perp = 0,45,90^\circ$, at $b=0.63$ (Top) and $b=0$ (Bottom). These signals are for a Jupiter-size ($R_{\rm p}/R_\star=0.1$) planet with a Saturn-like oblateness ($f_\perp=0.1$). The black curves are computed for stars with uniform surface brightness, and the blue solid curve is for a limb-darkened Sun-like star.
\label{fig:examples}}
\end{figure*}

The transit light curve of an oblate planet is described by the intersection between an ellipse and a circle. For the ingress and egress regions of the light curve, where the deviation from a spherical planet model light curve is greatest, we compute the intersection region using the quasi-Monte Carlo integration algorithm introduced by \citet{Carter:2010a}.
\footnote{The correct form of Equations (B5) and (B6) of \citet{Carter:2010a} should be $r=\sqrt{u+(1-u)(a_1/a_2)^2}$ and $\theta=(1-v)\theta_1+v\theta_2$, respectively.}
For the in-transit regions, where the planet is fully within the stellar disk, we replace the oblate planet signal with a spherical planet with the same cross-section area. Since $R_p \ll R_\star$, we assume the spatial limb darkening variation within the area covered by the planet ellipse is negligible. In fact, the in-transit signal induced by limb darkening is $<10$ ppm even for planets with substantially large $R_p/R_\star$ \citep{Carter:2010a}, which is insignificant for the purpose of model-fitting compared with the noise level ($\sim$500 ppm) of \emph{Kepler} short cadence light curves. By replacing the oblate planet with a spherical planet for the in-transit region, we are able to speed up the original algorithm introduced by \citet{Carter:2010a} by a factor of 10. We demonstrate in Figure~\ref{fig:examples} the theoretical oblateness-induced signal for a Jupiter-size planet in a 100-day orbit around a sun like star with uniform surface brightness (i.e., no limb darkening effect).
We assume the planet has a Saturn-like oblateness ($f_\perp=0.1$, we note that $f_{\rm saturn} = 0.098$). Signals from different projected planet spin obliquities $\theta_\perp$ ($0^{\circ}$, $45^{\circ}$, $90^{\circ}$) are plotted for comparison. We also add an example signal from the oblate planet transiting a limb-darkened stellar surface in Figure~\ref{fig:examples}, with the in-transit signal considered, to show the influence of the limb darkening effect.

As Figure~\ref{fig:examples} shows, the oblateness-induced signal is at least an order of magnitude greater over ingress and egress than the in-transit part. The theoretical maximum amplitude of the signal at the ingress/egress part can be estimated by
\begin{equation} \label{eq:amplitude}
  {\rm Signal}_{\rm max} \approx \frac{f_\perp}{2 \pi} \left(\frac{R_{\rm p}}{R_\star}\right)^2
  = 160\ {\rm ppm} \left(\frac{R_{\rm p}/R_\star}{0.1}\right)^2 \left( \frac{f_\perp}{0.1} \right), 
\end{equation}
which  can be achieved when the impact parameter satisfies the following condition:
\begin{equation}
  b = -\min [{\rm abs}(\sin{\theta_\perp}),{\rm abs}(\cos{\theta_\perp})].
\end{equation}
The derivation for this expression of the maximum achieved signal is shown in detail in Appendix~\ref{sec:appendix-1}. Although Equation~(\ref{eq:amplitude}) is useful in estimating the amplitude of oblateness-induced signal, one should keep in mind that the recovered signal from light curve modeling may have a smaller amplitude due to slight changes in those from other fitting parameters, especially when the oblateness-induced signal is mirror symmetric \citep{Barnes:2003}. Space-based observations, such as {\em Kepler}, can achieve photometric precision comparable to this signal amplitude \citep{Jenkins:2010,Gilliland:2010,Murphy:2012}, making possible the detection of planetary oblateness when multiple transits are observed.

It is also useful to have a comprehensive understanding of the amplitude of the oblateness signal for arbitrary projected obliquity angles, $\theta_\perp$, and impact parameters, $b_0$. To achieve that, we simulated theoretical light curves on a grid of $\theta_\perp$ and $b_0$ with fixed $f_\perp$ and $R_{\rm p}/R_\star$. For most of the ranges of $\theta_\perp$ and $b_0$ we consider, assuming a moderate projected oblateness $f_\perp$ ($\lesssim 0.2$), Equation (\ref{eq:amplitude}) can be written as
\begin{equation}
  {\rm Signal}_{\rm max} (R_{\rm p}/R_\star,f_\perp,\theta_\perp,b_0) 
  \approx \frac{f_\perp}{2 \pi} \left(\frac{R_{\rm p}}{R_\star}\right)^2 g(\theta_\perp,b_0) ,
\end{equation}
In Figure~\ref{fig:b0-sina} we show an example of ${\rm Signal}_{\rm max} (R_{\rm p}/R_\star,f_\perp,\theta_\perp,b_0)$ with $R_{\rm p}/R_\star=0.1$ and $f_\perp=0.1$. This figure tells us that the amplitude of oblateness signal reaches maximum at $\theta_\perp \sim 0^\circ$ for a system that is very close to edge-on, and at larger (absolute) projected obliquities as the system becomes more inclined. Particularly, for $b_0 \approx \pm 0.7$, the signal will have maximum amplitude at $\theta_\perp \approx \pm 45^\circ$ where the shape of the signal becomes the most asymmetric (see Figure~\ref{fig:examples}). This is why the detectability of oblateness is maximized at $b\approx0.7$, as determined numerically by \citet{Barnes:2003}. The values of $\theta_\perp$ where the maximum amplitude of signal is achieved are also the ones where the oblateness can be most tightly constrained. We will further demonstrate this point in the modeling of simulated signal and real data. 

If limb darkening is considered, the maximum amplitude decreases by a factor of $6(1-u_1-u_2)/(6-2u_1-u_2)$, where $u_1$ and $u_2$ are the quadratic limb darkening coefficients \citep{Claret:2000}. 

\subsection{Detectability in {\em Kepler} Data} \label{sec:detectability}

As discussed in Section \S2.1, for a Jupiter-size planet ($R_{\rm p}/R_\star = 0.1$) with Saturn-like oblateness ($f=0.1$), the maximum signal amplitude due to oblateness is $\sim$160 ppm. 

The signal-to-noise ratio (SNR) for a single transit in the white noise limit can be estimated by 
\begin{equation} \label{eq:snr}
{\rm SNR} = \left(\frac{f_\perp}{0.1}\right) \left(\frac{R_p/R_*}{0.1}\right)^2 \left(\frac{\rm ootv}{160{\rm ppm}}\right)^{-1} \sqrt{2N_{\rm p}},
\end{equation}
in which, ${\rm ootv}$ is the out-of-transit variation amplitude (per point), $N_{\rm p}$ denotes the number of observations during the ingress/egress time. The duration of the ingress/egress time $\tau$ can be estimated by
\begin{equation} \label{eq:duration}
\tau \approx \frac{2R_{\rm p}}{v_{\rm orb}} = 50.4\ {\rm mins} \left( \frac{R_{\rm p}/R_\star}{0.1} \right) \left(\frac{R_\star}{R_\odot}\right) \left(\frac{P_{\rm orb}}{100\ \rm days}\right)^{1/3} \left(\frac{M_\star}{M_\odot}\right)^{-1/3} \ .
\end{equation}

For a 12$^{\rm m}$ magnitude star, the {\em Kepler} long cadence data can typically achieve a photometric precision of $100 \,{\rm ppm}$. Assuming white noise, the equivalent out-of-transit scattering (${\rm ootv}$) of the corresponding short cadence data is $\sim 550$\,ppm, which corresponds to SNR$\sim 3$. We note that this SNR estimation is almost exact for short cadence data, with the oblateness-induced signal sufficiently mapped out by the 1 min cadence. Transits having only long cadence data will typically have a much shallower signal than the theoretical value, given the signal duration is roughly comparable to the long integration time (30 min). For example, for a Jupiter-size planet in a 15-day orbit, the ingress/egress duration is $\sim$30 min, and thus the oblateness signal is strongly suppressed in the long cadence data. For this reason, we do not consider the long cadence data in the present work.

With {\em Kepler} light curves from Q1-Q16 ($\sim$ 1400 day), we expect a planet with a 100-day orbital period and a Saturn-like oblateness to be detected with a SNR~$\sim 11$ in the white noise limit. In the red noise limit, SNR would scale with the number of transits instead of the number of data points observed, which would reduce the expected SNR severely.

\section{Target Selection and Light Curve Trend Removal} \label{sec:data}

We select {\em Kepler} planet candidates from Q1-Q16 \citep{Batalha:2013,Burke:2014}  with the following criteria: 

a) candidates with radius smaller than 2.0 $R_J$, to avoid eclipsing binaries; 

b) orbital periods longer than 15 days, such that the spin-orbit synchronization timescale of the planet is greater than 250 Myr, assuming a Jupiter-mass planet around a solar-mass star \citep{Carter:2010b}; 

c) the transit is not grazing ($b<0.8$), such that the transit system parameters are well constrained;  

d) expected SNR for a planet with a Saturn like oblateness is higher than 0.5 for a single transit.

We use Equation \ref{eq:snr} to compute the SNR per transit for each candidate. The ootv value is the point-to-point scatter of the light curve, measured over the regions on either side of the transit event, with width of half transit duration. The actual expected SNR will be multiplied by the square root of the available number of transits from {\em Kepler} data. In addition, we did not select those systems with strong transit timing variations, for it leads to more complexity in the modelling. 11 candidates pass the above threshold, among which, only 3 have short cadence data. We select these three targets as examples for the modeling. They are KOI 686.01, KOI 197.01 and KOI 423.01 (Kepler 39b). We note that these three targets together are not a statistically meaningful sample since our selection criteria are somewhat arbitrary. We also modeled HAT-P-7b (KOI 2.01), despite the fact that it is a hot Jupiter and outside of our period selection range, since it will provide a direct comparison to the modeling of HD~189733b in \citet{Carter:2010a}. 

To characterize our detection sensitivity to the oblate planet signals, we perform a signal injection and recovery exercise using the light curves of KOI-368. The 11.3 magnitude star is photometrically quiet, with near-minimum ootv of a bright star in {\em Kepler} band. The KOI-368 system, with 110-day period, resembles an ideal system, with near-optimal oblateness SNR detectability. We chose not to fit for oblateness in the actual transits of KOI-368.01, since 1) KOI-368.01 is an M-dwarf, not a planet, and 2) the modeling would require a simultaneous fit of the stellar gravity darkening effect \citep{ZhouHuang:2013} and the planet oblateness effect, which is outside the scope of this study. 

We detrended all the available public {\em Kepler} short cadence light curves of the above targets for our analysis. To remove the stellar variability, we use the raw flux (Simple Aperture Photometry, $\rm SAP\_FLUX$) obtained from the MAST archive
\footnote{http://archive.stsci.edu/kepler/data$\_$search/search.php}, with the 
out-of-transit variations corrected by the following steps from \citet{Huang:2013}:

a) removal of bad data points;

b) correction of systematics due to various phenomena of the space craft, such as safe modes and tweaks;

c) we use either a set of cosine functions with minimum period of 3 times the transit width \citep{Kipping:2013b} to fit over the out-of-transit regions and correct the trend in the light curves if there are high-amplitude, short term (1 day - 20 days) variations present in the light curves. Otherwise, we use a 7th order polynomial as our fitting function for the out-of-transit variations. 

\section{Light Curve Fitting and results}

In this section, we first perform a signal injection and recovery exercise with simulated transits of KOI 368.01 to demonstrate the modeling and fitting process. We then present the modeling of the four selected systems: HAT-P-7b (KOI 2.01), KOI 686.01, KOI 197.01, and KOI 423.01 (Kepler 39b). We first fit a standard \citet{MandelAgol:2002} transit model to the light curves, with the free parameters being the orbital period $P$, transit epoch $T_0$, planet-to-star radius ratio $R_p/R_\star$, normalized orbital semi-major axis $(R_p+R_\star) / a$, line-of-sight inclination $i$, and quadratic limb darkening parameters $q_1$ and $q_2$ parameterized according to \citet{Kipping:2013}
\begin{equation} \label{eq:ld-coeff}
q_1 \equiv (u_1+u_2)^2;\ q_2 \equiv \frac{u_1}{2(u_1+u_2)} .
\end{equation}
For the initial positions of the minimization, we use system parameters from the cumulative {\em Kepler} planetary candidate table of the \emph{NASA Exoplanet  Archive}
\footnote{\url{http://exoplanetarchive.ipac.caltech.edu/}}
and limb darkening coefficients from \citet{Sing:2010}. We then introduce the oblateness parameters $f_\perp$ and $\theta_\perp$, and rerun the minimization using the oblateness model, with the best fit values from the standard model fit as starting points. To avoid artifacts from boundary conditions, we allow $f_\perp$ to vary from -0.6 to 0.6, taking the absolute value and then folding the negative and positive $f_\perp$ steps together in the final analysis. Similarly, we also allow periodic boundary conditions for $\theta_\perp$ between $-90^\circ$ and $90^\circ$. 

For each minimization, we perform a Markov chain Monte Carlo (MCMC) analysis to find the best fit parameters and the associated uncertainties using the \texttt{emcee} ensemble sampler \citep{ForemanMackey:2013}. In each fitting routine, we employ 150 walkers each with 4000 iterations. We chose the number of walkers to match the suggested acceptance rate by \texttt{emcee}. After the MCMC routine finishes, we flatten the whole chain and discard the initial steps ($\leq 25\%$) where the sampling has not burnt-in. We confirmed the convergence of the chain by splitting the whole chain, after the burn-in steps were discarded, into two parts and comparing the posteriors from them.

We give the comparison between the {\em Kepler} parameters and our best-fit parameters for each planet. The same as \citet{Carter:2010a}, for most parameters we report the median of the posterior probability distribution, along with error bars defined by the $16\%$ and $84\%$ levels of the cumulative distribution. For the projected oblateness $f_\perp$, we report the $68\%$ confidence upper limit if the peak of $f_\perp$ has no obvious deviation from zero. Otherwise we report the median value, together with the $16\%$ and $84\%$ levels of the cumulative distribution as the error bars. We do not report constraints on the projected obliquity since it is weakly constrained, as \citet{Carter:2010a} has shown.

For each system, we also inject null signal transit light curves, resembling that of a spherical planet with the same system parameters as the target system, into different out-of-transit sections of the SAP light curve. The injected transits are shifted in time to the same epoches as the true transits. The simulated null-case transits are then processed in the same way as the observed transits.

\subsection{Injection and Recovery Simulations with KOI 368.01} \label{subsec:368}

\begin{figure}
\centering
\epsscale{1.1}
\plotone{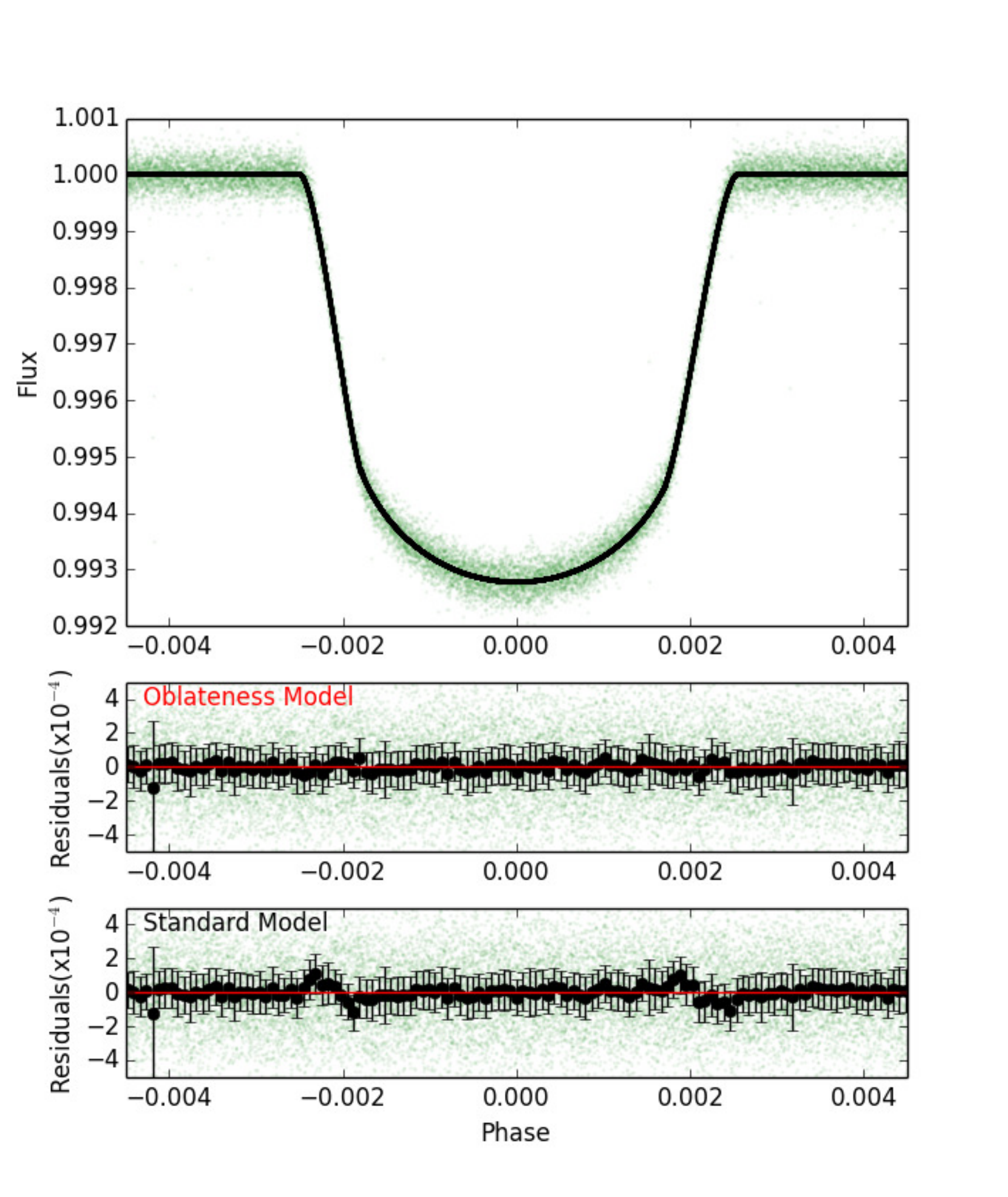}
\caption{The simulated transit of an oblate planet, with the same system parameters as that of KOI-368.01, with oblateness of $f_\perp = 0.1$ and obliquity $\theta_\perp=45^\circ$, into the out-of-transit light curve of KOI-368 (Top). The fit residuals with 12 transits to the best-fit oblate planet model (middle panel) and the best-fit standard transit model (bottom panel) are plotted. We binned the data in the phase space with a 150 points moving window and show the median and scattering in the black dots.
\label{fig:fake-lc}}
\end{figure}

\begin{figure*}
\centering
\epsscale{0.8}
\plottwo{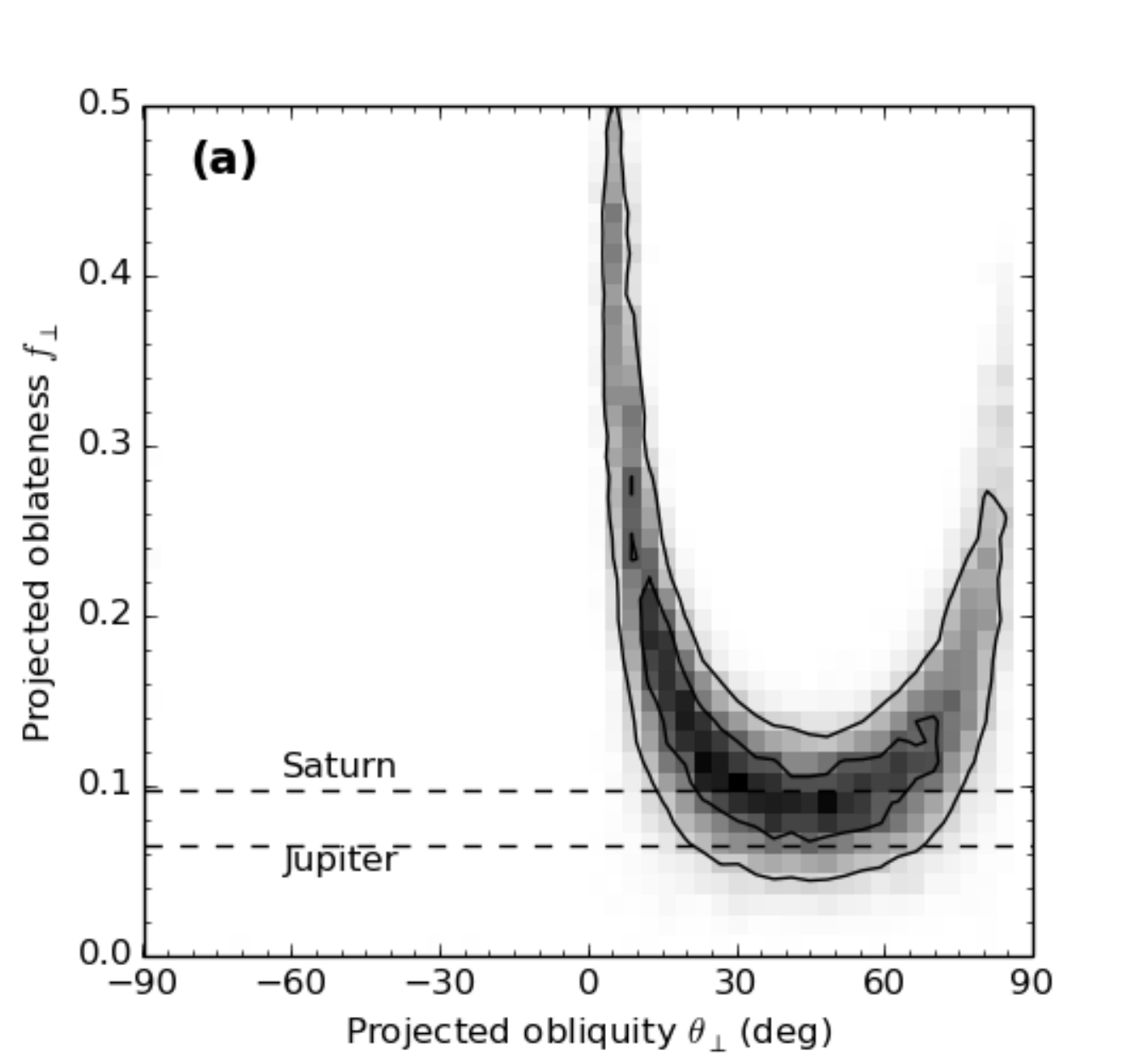}{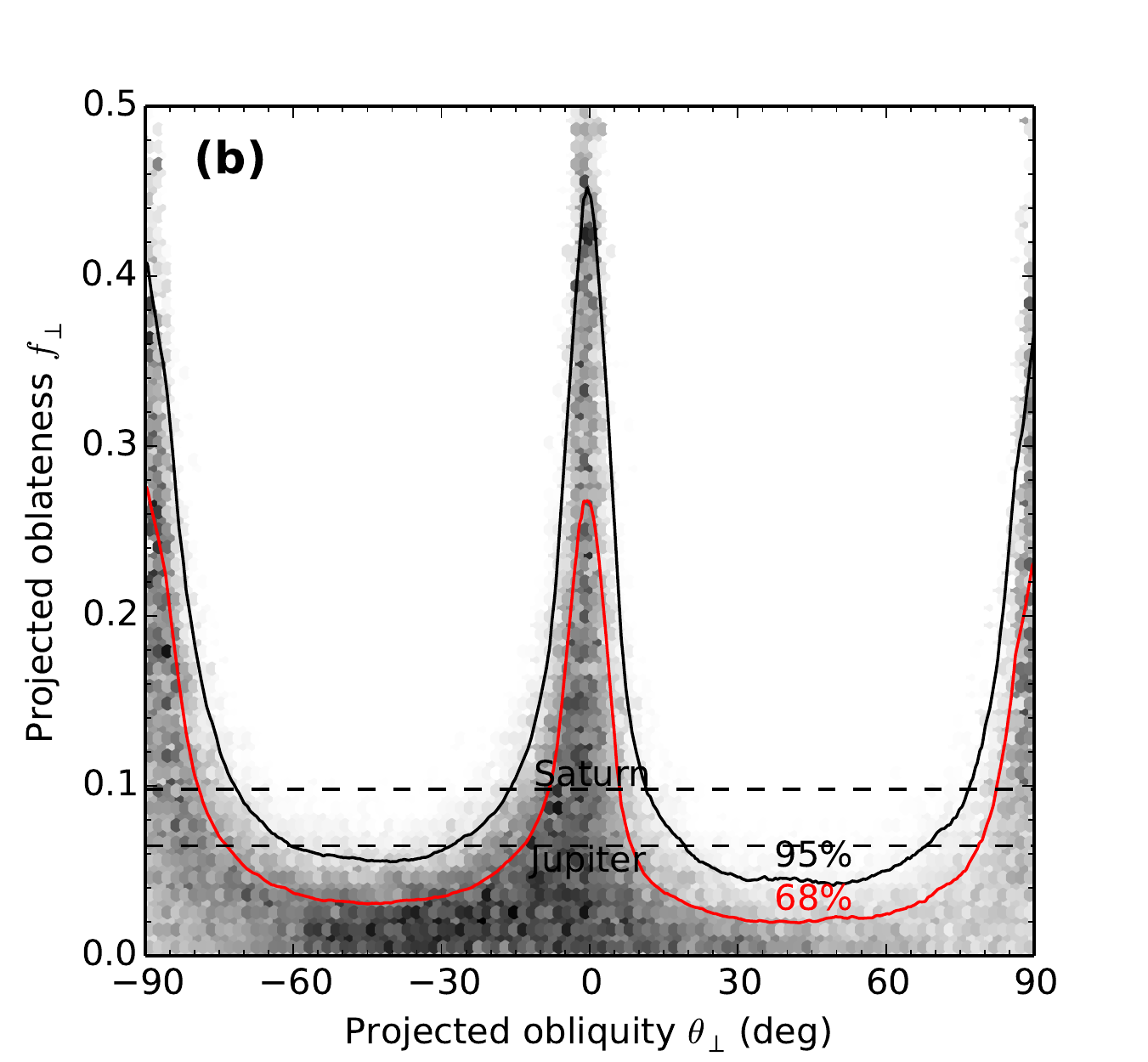}
\plottwo{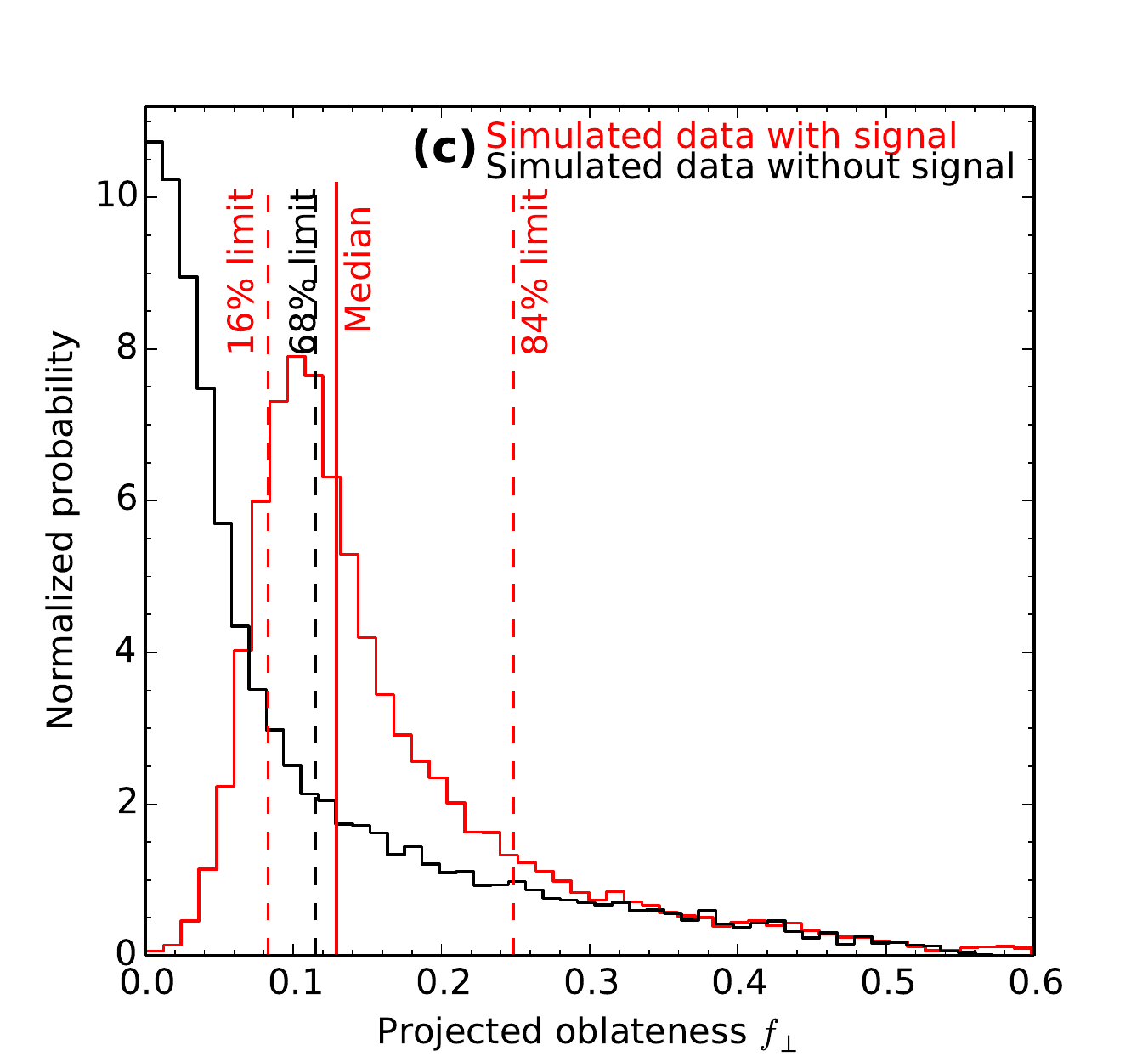}{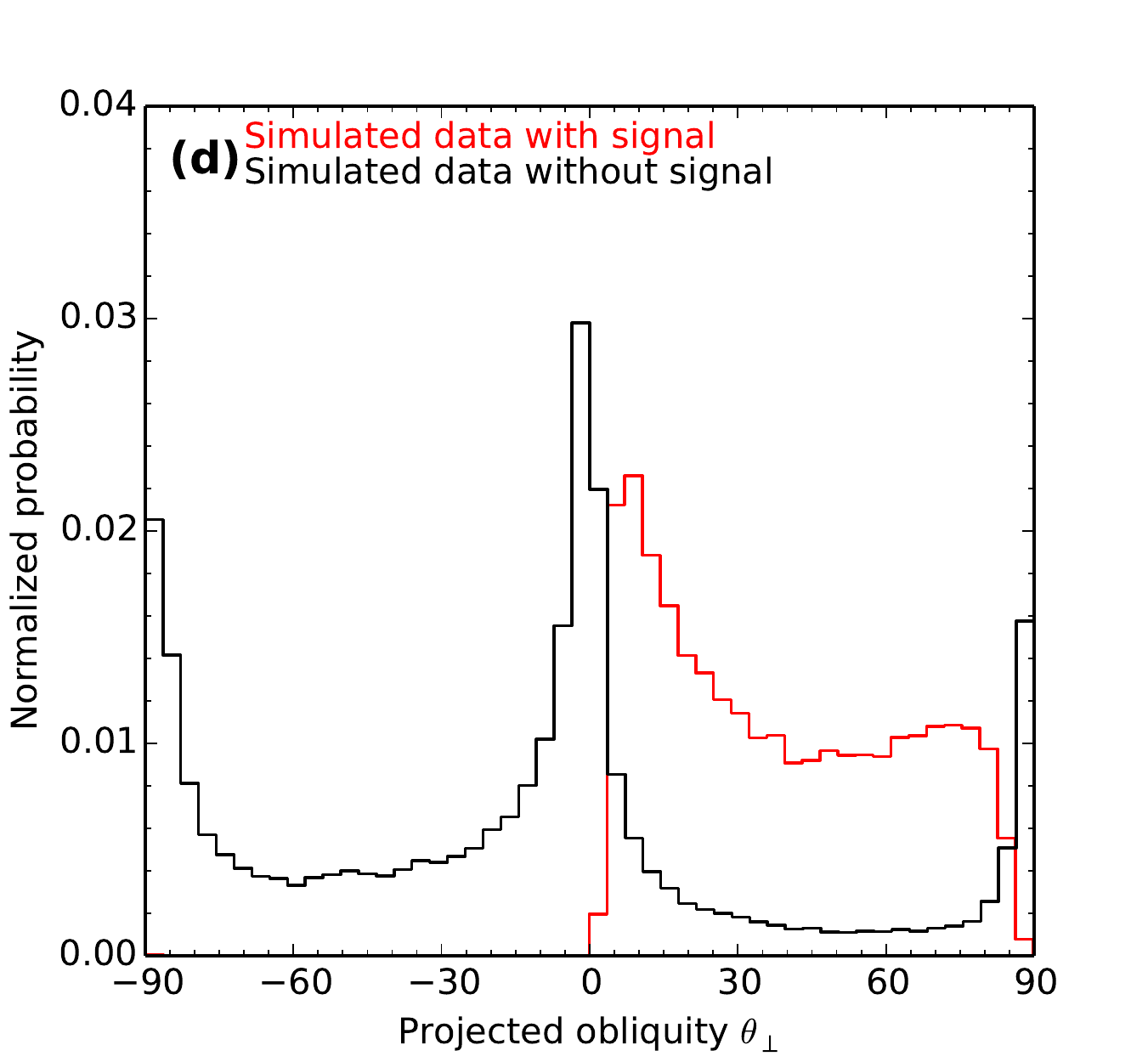}
\caption{\textit{Panel (a)}: 2D posterior distributions for the projected oblateness $f_\perp=0.1$ and obliquity $\theta_\perp=45^\circ$ of a single simulated KOI 368.01 oblate planet transit. \textit{Panel (b)}: Posterior distributions for the simulated KOI 368.01 standard Mandel-Agol transit (null case). \textit{Panel (c) and (d)}: Marginalized distributions of $f_\perp$ and $\theta_\perp$ for simulated cases with (red) and without (black) the oblateness signal. 
\label{fig:fake-posterior}}
\end{figure*}

\begin{figure*}
\centering
\epsscale{1.0}
\plotone{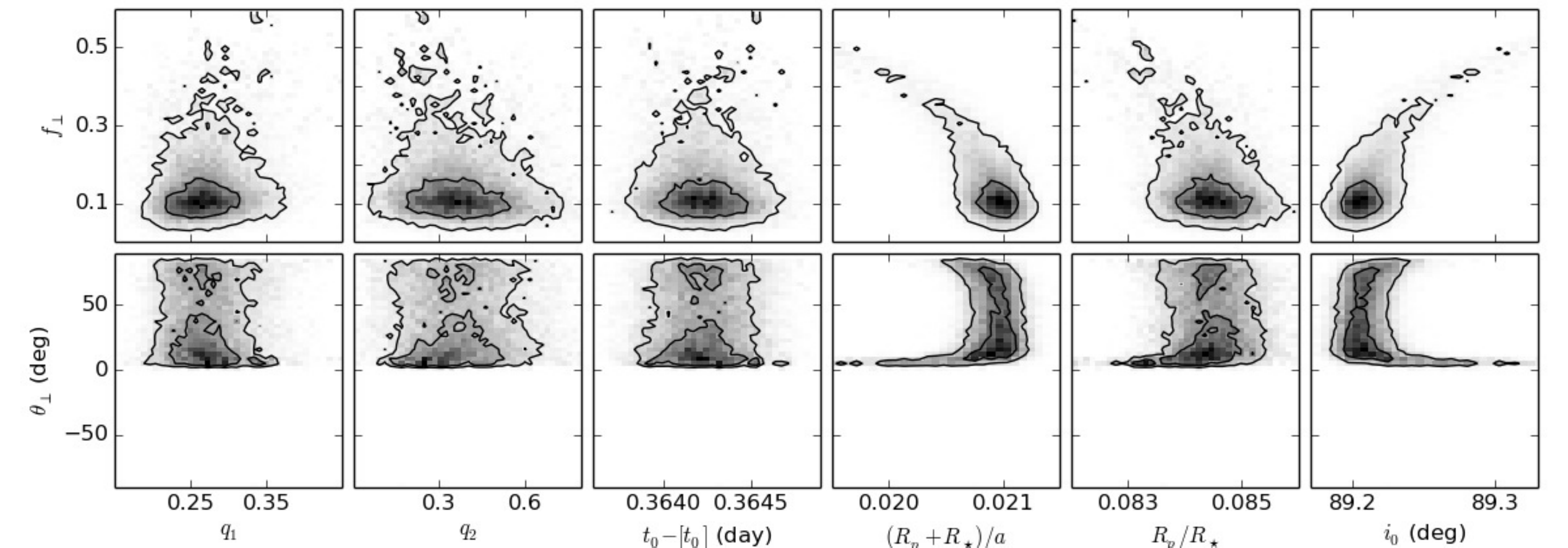}
\caption{Correlations between $f_\perp$, $\theta_\perp$ and other transit parameters, as derived from the single simulated KOI-368.01 oblate planet transit.
\label{fig:correlations-368}}
\end{figure*}

\begin{figure*}
\centering
\epsscale{0.8}
\plottwo{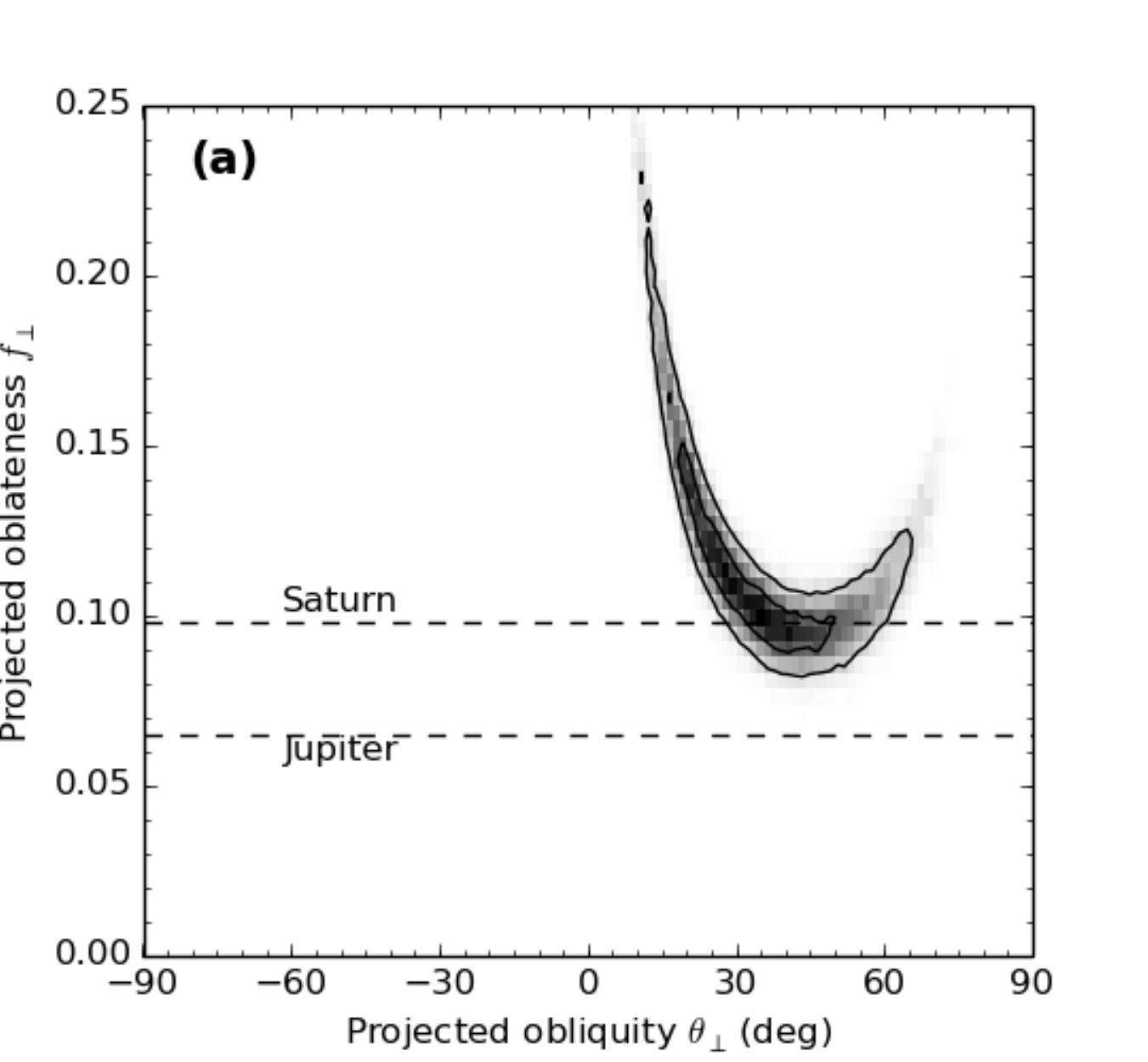}{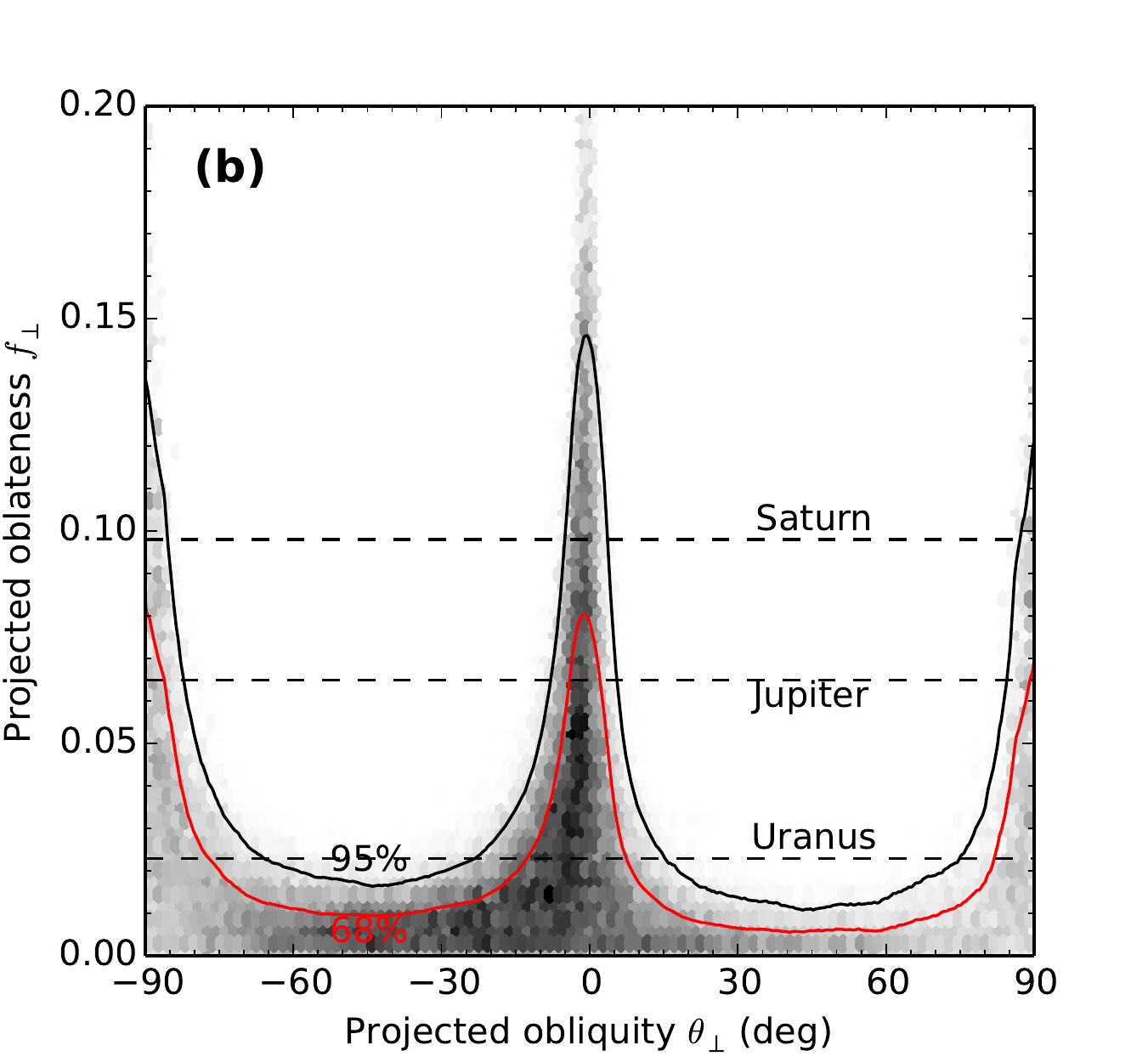}
\plottwo{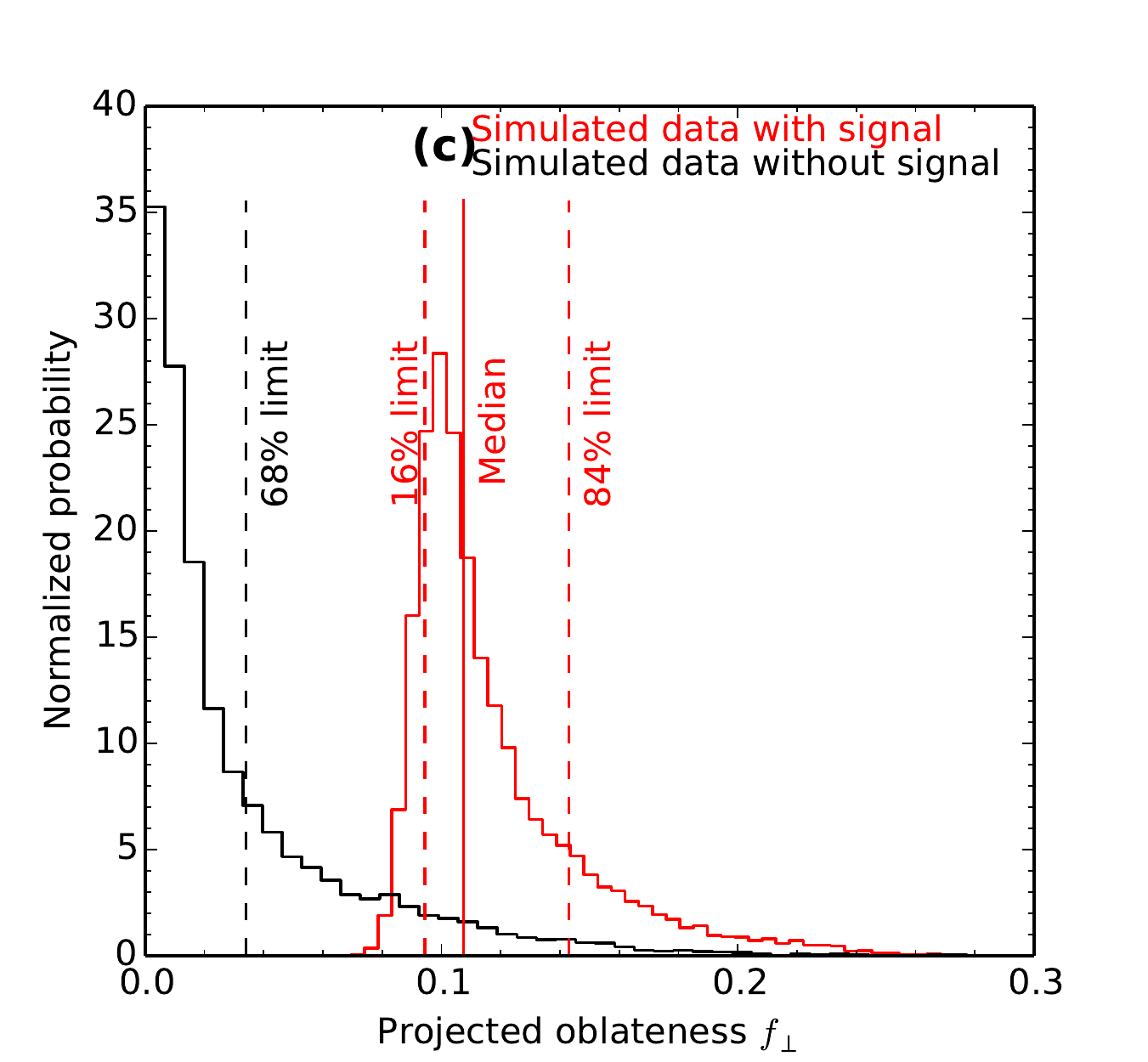}{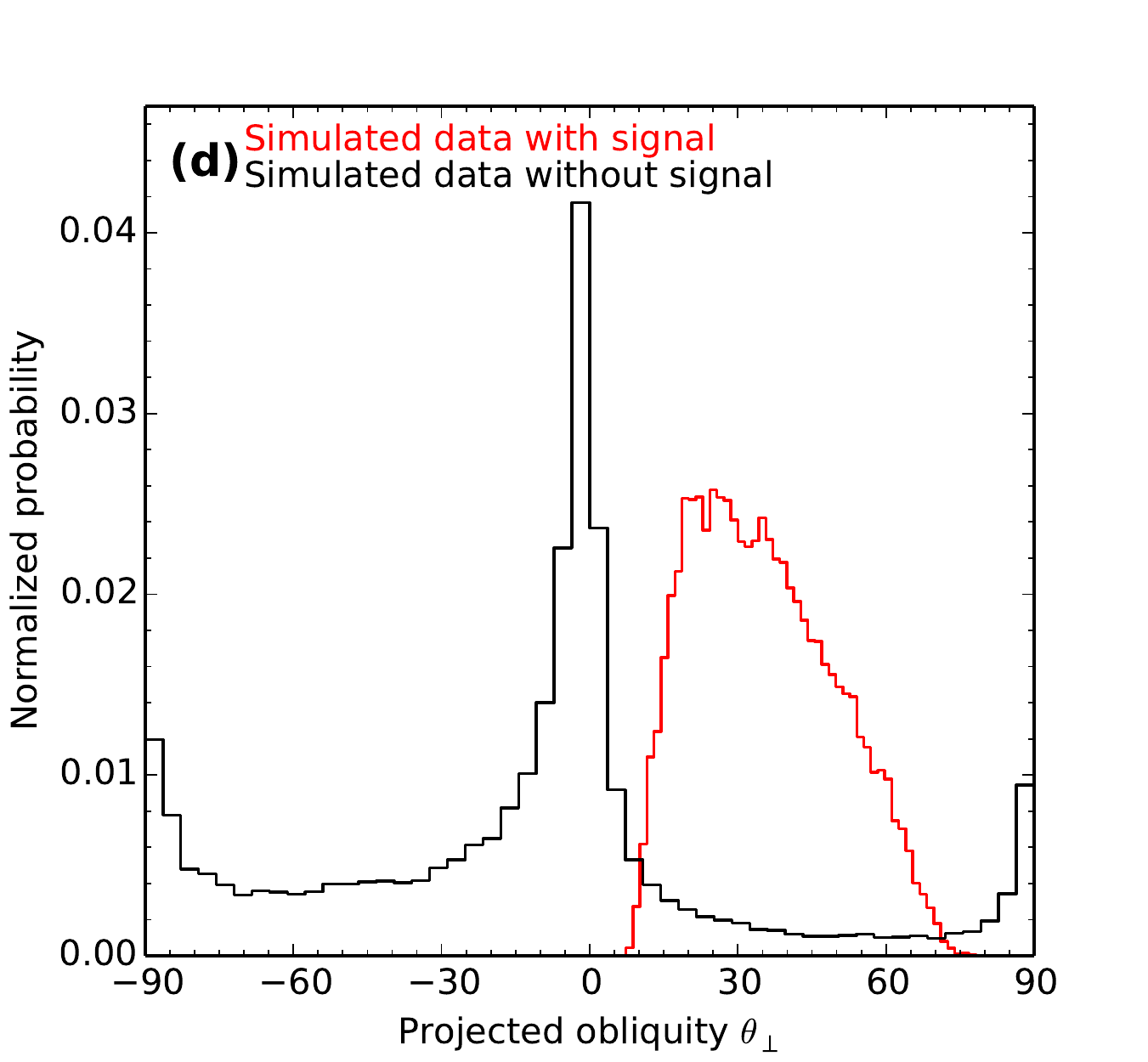}
\caption{The posterior distributions as recovered from 12 injected
  KOI-368 oblate planet transits. Figure captions as per Figure~\ref{fig:fake-posterior}.
\label{fig:fake-posterior-12}}
\end{figure*}

To demonstrate the detectability of the oblateness signal, we perform a signal injection and recovery exercise using out-of-transit light curve segments from the KOI-368 system. KOI-368 is a photometrically quiet A-dwarf hosting a transiting M-dwarf with $P=110.32$ day, $R_p/R_*=0.084$, $i=89.36$ ($b=0.64$), and limb darkening coefficients $u_1=0.23$, $u_2=0.29$ \citep{ZhouHuang:2013}. The KOI-368 light curves and system properties resemble the transits of an optimal case long-period planet candidate. We inject 12 transits of an oblate planet with the same system parameters as KOI-368.01, and $f_{\perp}=0.1$ and $\theta_{\perp}=45^\circ$, onto random segments of the SAP short cadence light curves of KOI-368. The injected transit yields an oblateness signal with amplitude 80 ppm. In contrast, the out-of-transit variation amplitude of the light curves is 300 ppm. 

The simulated light curve is then processed by our standard detrending pipeline (see Section \S\ref{sec:data}) and then fitted with the MCMC method. For comparison, we also report the results with standard Mandel-Agol transits injected in the same segments and detrended in the same way. In each case, the simulated data is first fitted by a standard transit model and then an oblate planet model. To demonstrate the least and most optimal cases, we present the injection and recovery for the case of a single transit, and the case of all 12 full transits for the KOI 368 system. The light curves and the residuals to the standard and oblate planet models (for the 12 transits case) are shown in Figure~\ref{fig:fake-lc}. 

The signal is detected at a significant confidence level in our oblate planet model fitting. The original injected oblateness $f_\perp$ is recovered in the single transit case. The left panels in Figure~\ref{fig:fake-posterior} show the posterior distributions from the fitting of a single injected short cadence transit that contains the oblateness signal.  Although the projected obliquity suffers the degeneracy between $f_\perp$ and $\theta_\perp$ and is therefore not well recovered, the asymmetry between $\theta_\perp>0$ and $\theta_\perp<0$ implies the true value should be positive. 

For comparison, the posteriors of the null case are shown on the right panels of Figure~\ref{fig:fake-posterior}. The null case here is also an example that can be used to demonstrate the ability of constraining the oblateness with {\em Kepler} data. From the fitting of a single injected transit, an oblateness as large as that of Saturn can be ruled out if the planet has moderate obliquity. However, the constraint we can place at $\theta_\perp=0, \pm 90^\circ$ is very weak, since for this case $b=0.64$, the oblateness signal can still be very small even for large oblateness, as is shown in Figure~\ref{fig:b0-sina}.

\begin{deluxetable*}{cccccc}
\tablecaption{Signal injection and recovery parameters for KOI-368 (12 transits)
\label{tab:k368-fitting}}
\tablehead{Parameters & Injected (Oblate Model) & Standard Model & Oblate Model & Injected (Standard Model) & Oblate Model}
\startdata
\input{k368-fitting.dat}
\enddata
\end{deluxetable*}

The correlations between oblateness parameters $f_\perp$, $\theta_\perp$ and other transit parameters, derived from the MCMC fitting of a single injected oblate planet transit, are shown in Figure~\ref{fig:correlations-368}. $f_\perp$ appears to be correlated with $(R_{\rm p}+R_\star)/a$, $R_{\rm p}/R_\star$ and $i_0$, and shows long tails toward large oblateness in each plot, since $\theta_\perp$ is only weakly constrained. These correlations suggest that one might get systematic biases on the planetary parameters if an oblate planet is fit with a spherical model. It is worthwhile noting that the $f_\perp$ is hardly correlated with the limb darkening coefficients $q_1$ and $q_2$.

When the number of transits is increased to 12, the injected signal is recovered with significantly higher confidence level, as is shown in Figure~\ref{fig:fake-posterior-12}. We can see the oblateness signal visually in the residual (from all 12 transits) of the fitting using only the standard Mandel-Agol model in Figure~\ref{fig:fake-lc}. For the null case, we can successfully rule out an oblateness as small as that of Uranus for most $\theta_\perp$. The injected parameters, and the recovered parameters from standard model and oblateness model are shown in Table~\ref{tab:k368-fitting}.

\subsection{HAT-P-7b (KOI 2.01)}

\begin{deluxetable}{cc}
\tablecaption{Physical parameters about the host star HAT-P-7 (KOI 2.01, KIC 10666592).
\label{tab:k2-host}}
\tablehead{Parameters & Values}
\startdata
\input{k2-physical.dat}
\enddata
\tablecomments{$^a$ Adopted from \citet{Pal:2008}.}
\end{deluxetable}

\begin{deluxetable*}{cccc}
\tablecaption{Fitting parameters of HAT-P-7b.
\label{tab:k2-fitting}}
\tablehead{Parameters & Kepler & Standard model & Oblate model}
\startdata
\input{k2-fittings.dat}
\enddata
\tablecomments{
\begin{description}
    \item[$^{\rm a}$] $q_1$ and $q_2$ are calculated using Equations~\ref{eq:ld-coeff} for the ``\emph{Kepler} model''.
\item[$^{\rm b}$] Adopted from \citep{VE:2013}.
\end{description}}
\end{deluxetable*}

\begin{figure*}
\centering
\epsscale{0.6}
\plotone{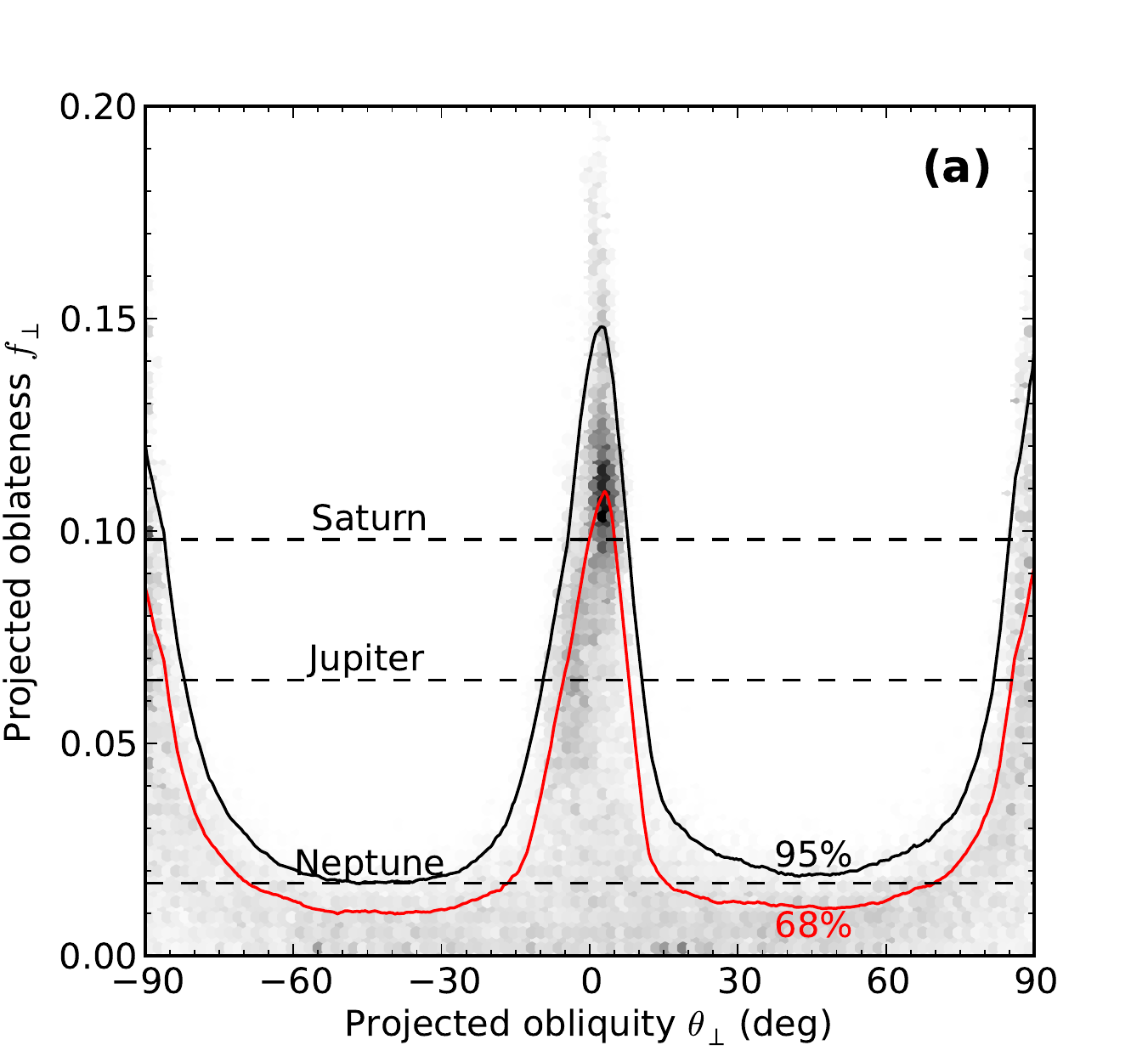}

\epsscale{0.9}
\plottwo{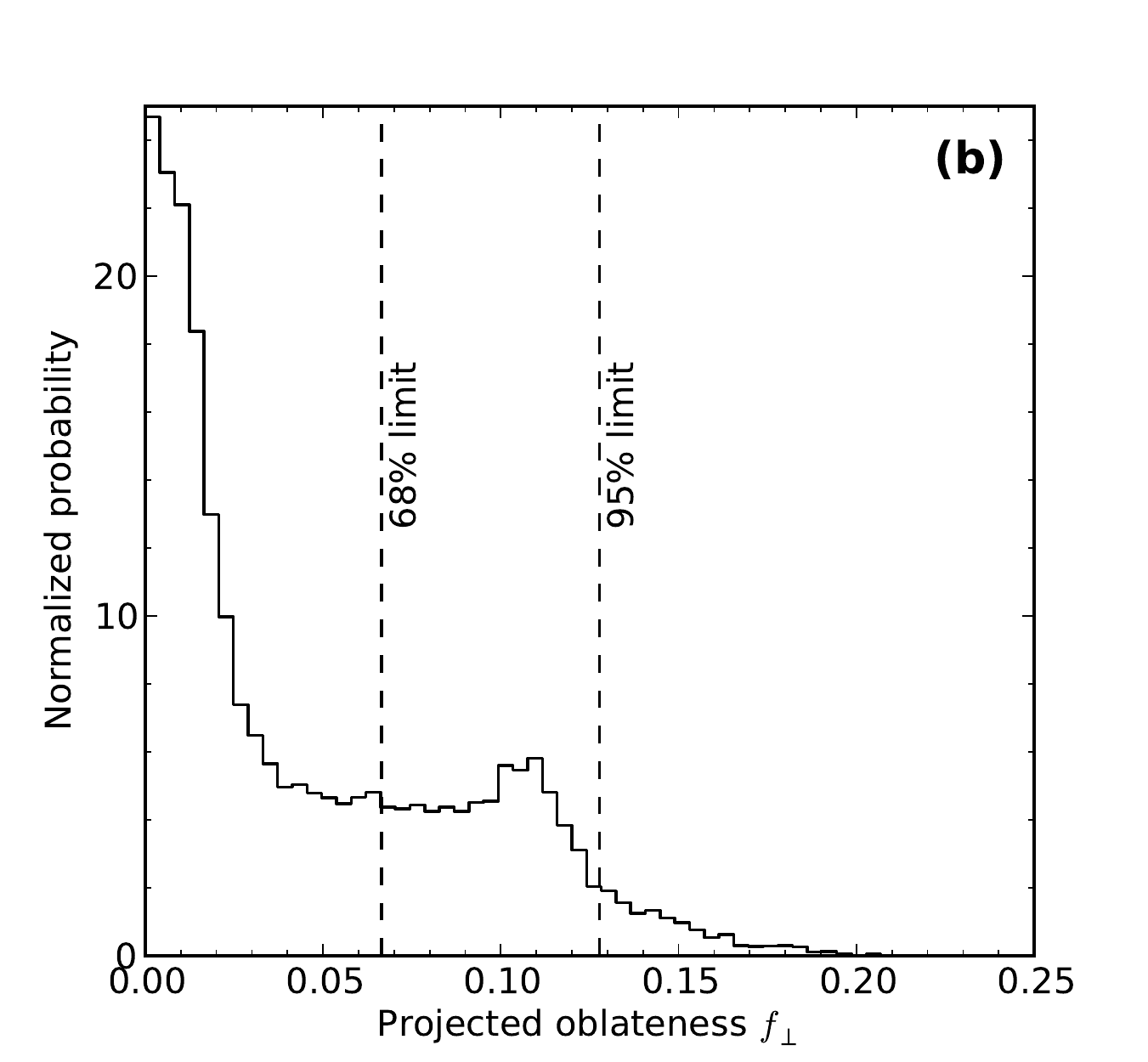}{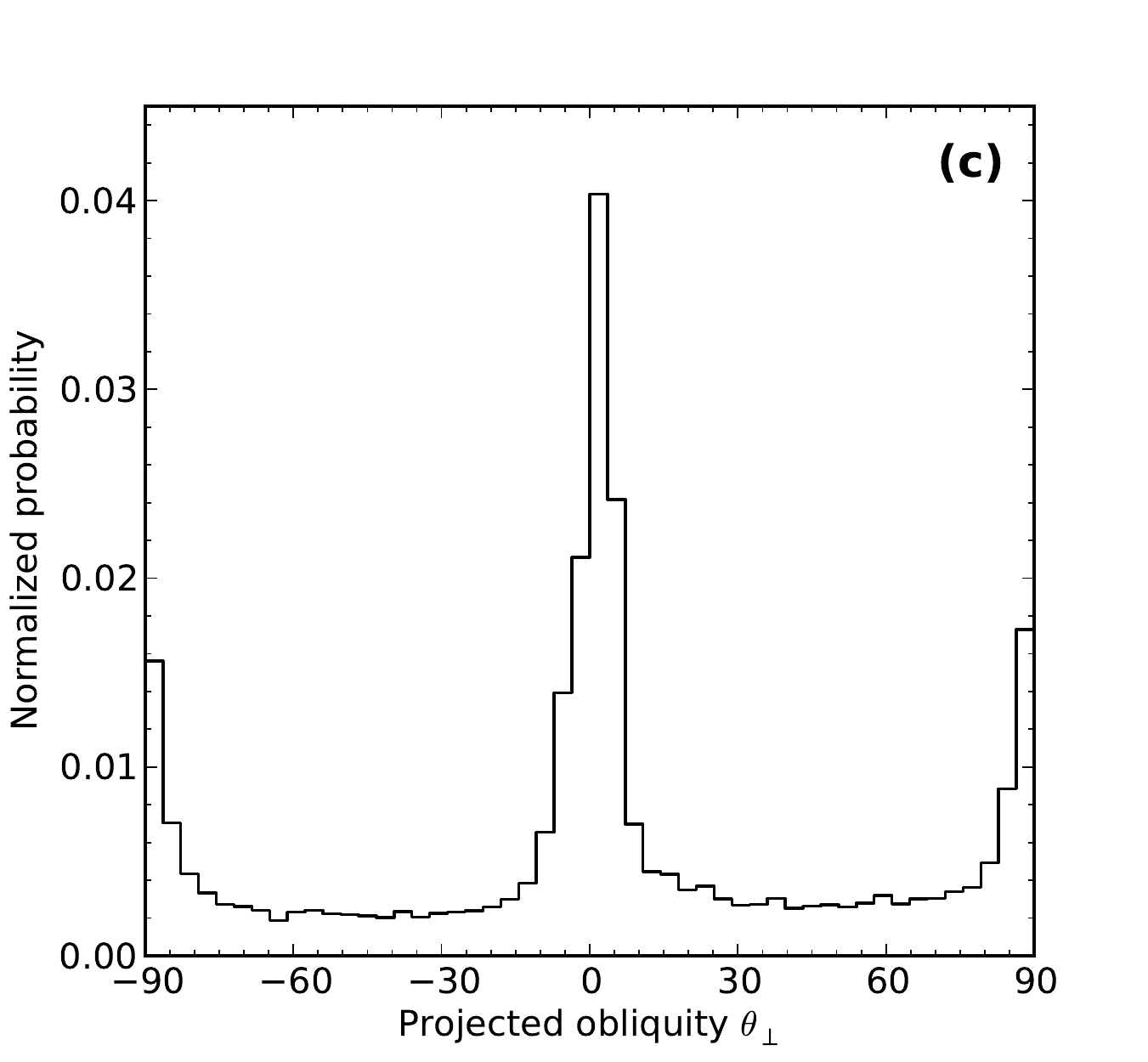}
\caption{Posterior distributions for the projected oblateness $f_\perp$ and obliquity $\theta_\perp$ for HAT-P-7b. a) The 2D posterior of $f_\perp$ and $\theta_\perp$ are plotted. The true oblateness of Saturn, Jupiter, and Neptune are marked for reference. The marginalized 1D posteriors for $f_\perp$ and $\theta_\perp$ are also plotted in panels b) and c) respectively.
\label{fig:hatp7-posterior}}
\end{figure*}

HAT-P-7b orbits around a slightly evolved F star in a 2.2-day orbit \citep{Pal:2008}. A planet in such a close-in orbit is believed to be tidally locked, and therefore HAT-P-7b should not exhibit rotation-induced oblateness. We include this planet in our sample for a number of reasons. First, this planet has been observed with short cadence mode all through {\em Kepler}'s operation time. There are in total 591 full transits of HAT-P-7b within Q1 -- Q16 short cadence data. The 1min cadence out-of-transit variation is 234 ppm. This will allow us to directly compare our modeling result with that of HD~189733b, the only other hot-Jupiter with its oblateness constrained by observations \citep{Carter:2010a}. Furthermore, there are various other photometric effects present in the HAT-P-7 light curves, such as the intrinsic stellar variability, planetary phase variations, stellar ellipsoidal variations \citep{Borucki:2009}, and possibly planet induced gravity darkening effects \citep{Morris:2013}. An analysis of HAT-P-7b can also help us understand how these effects affect our oblateness model fitting procedure. We list the physical parameters of the host star in Table~\ref{tab:k2-host}.

Fitting all 591 short cadence transits is extremely computationally expensive. It may also produce questionable results given the systematic transit depth variations between different quarters \citep{VE:2013}. We therefore fit the light curve of HAT-P-7b in three groups of 100 consecutive orbits each. The MCMC chains from the three groups are combined afterwards to arrive at the global best fit parameters. Examining the individual group results will also allow us to better judge the effect of stellar variability and red-noise on the consistency of our results. Our best-fit parameters for both standard and oblateness fittings, together with the initial parameters from {\em Kepler}, are listed in Table~\ref{tab:k2-fitting}. We show in Figure~\ref{fig:hatp7-posterior} the upper limits of $f_\perp$ for different $\theta_\perp$. The overall 68\% upper limit on the oblateness is 0.067. The oblateness is only weakly constrained around $\theta_\perp=0$ and $\pm 90^\circ$, which results in the small bump around $0.1$ in the posterior distribution of $f_\perp$. However, if the planet is moderately tilted, an oblateness as large as that of Neptune (0.017) can be ruled out with 95\% confidence. This constraint is comparable to that placed on HD189733b by \citet{Carter:2010a}, who ruled out a Uranus-like oblateness (0.023) (95\% confidence) for most of the projected obliquity angles and placed an overall 95\% upper limit of 0.058 on the planet. 

\subsection{KOI 686.01}

\begin{deluxetable}{cc}
\tablecaption{Physical parameters about the host star KOI 686 (KIC 7906882).
\label{tab:k686-host}}
\tablehead{Parameters & Values}
\startdata
\input{k686-physical.dat}
\enddata
\end{deluxetable}

\begin{deluxetable*}{cccc}
\tablecaption{Fitting parameters of KOI-686.
\label{tab:k686-fitting}}
\tablehead{Parameters & Kepler & Standard model & Oblate model}
\startdata
\input{k686-fittings.dat}
\enddata
\end{deluxetable*}

\begin{figure*}
\centering
\epsscale{1.0}
\plottwo{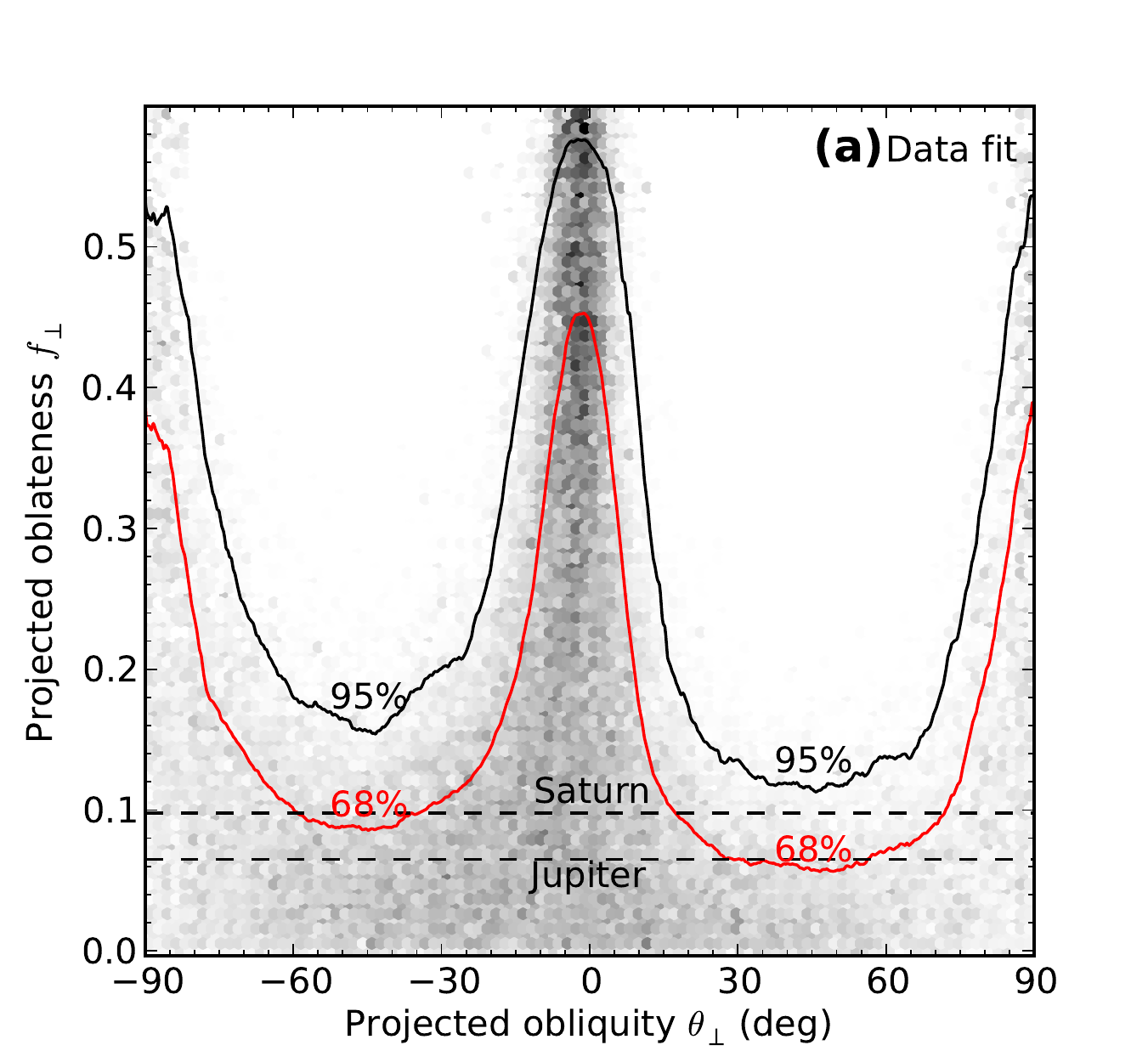}{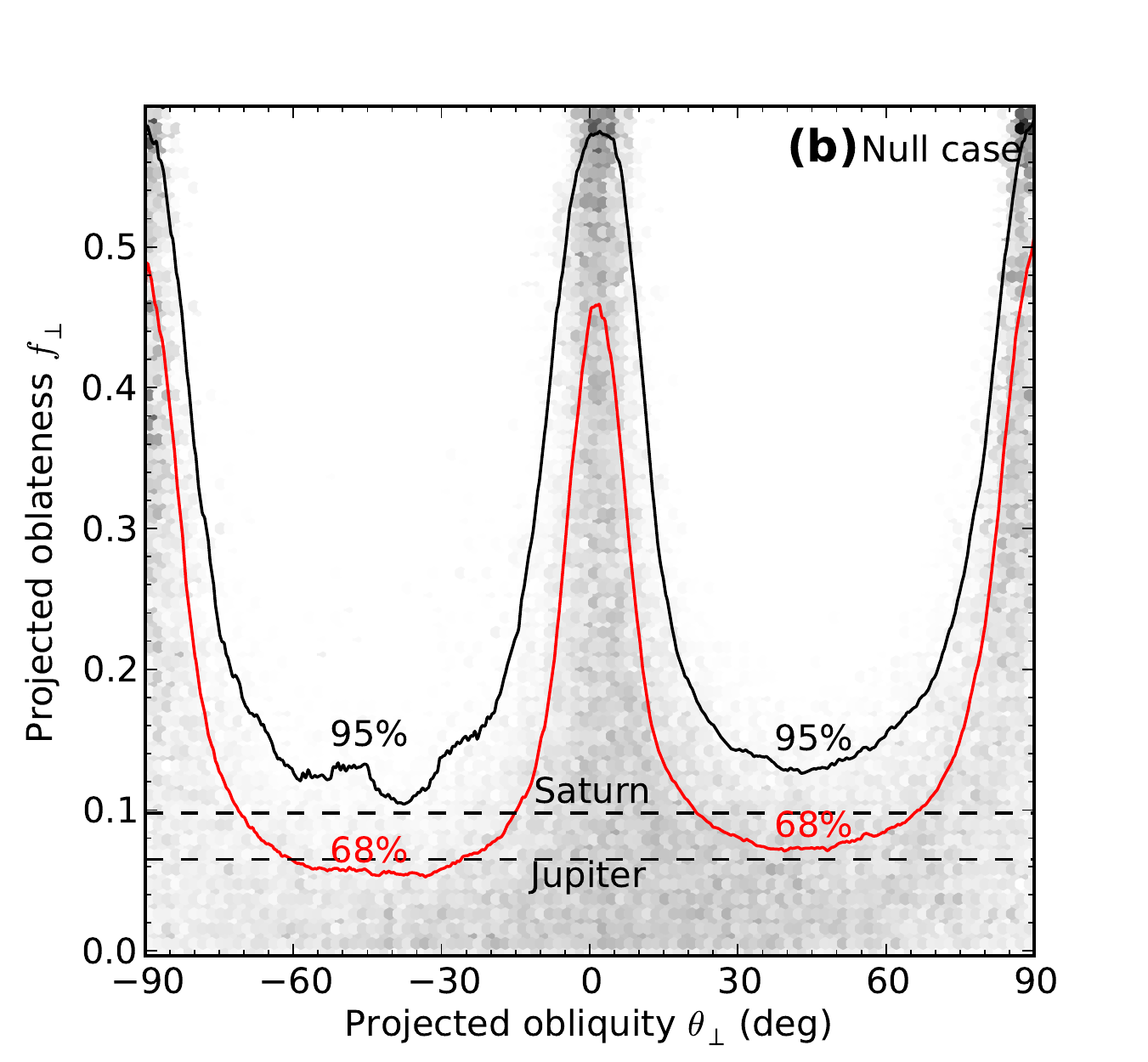}
\plottwo{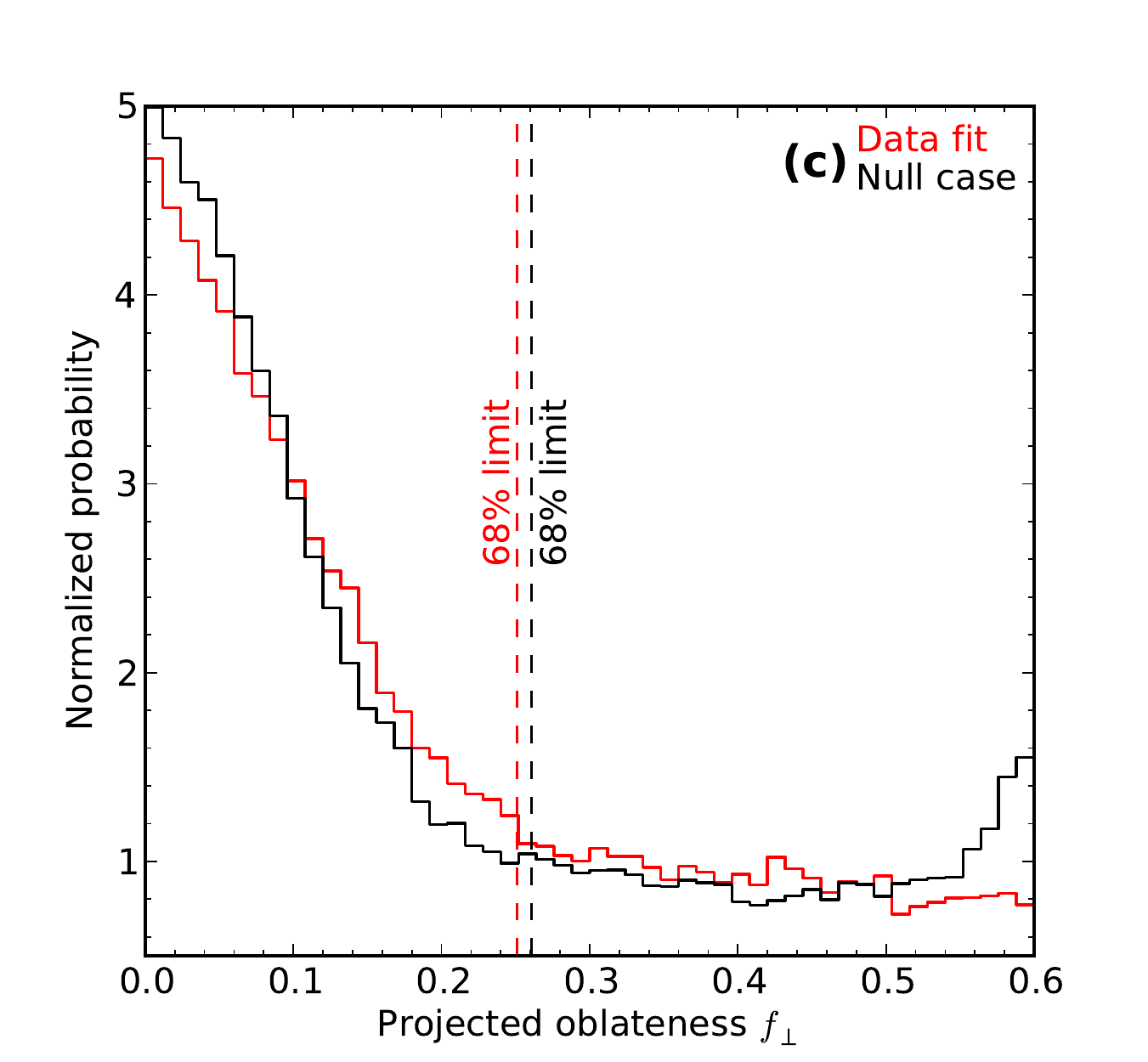}{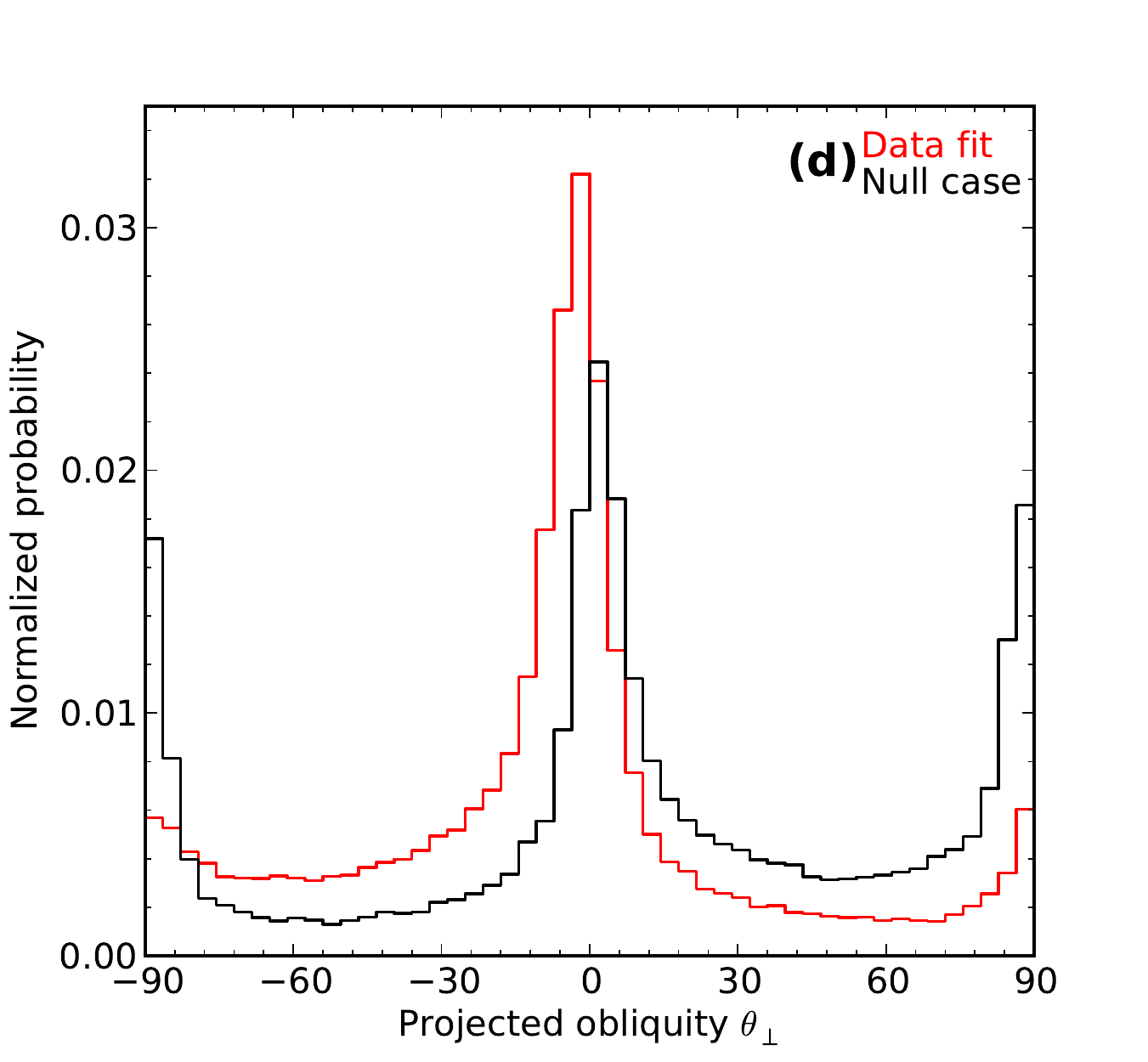}
\caption{The posterior distributions derived from the MCMC fittings of KOI-686.01. See caption of Figure~\ref{fig:hatp7-posterior} for details. The only difference from the HAT-P-7b case is that we also show the results from ``null case'' test. See the text for more details.
\label{fig:k686-posterior}}
\end{figure*}

KOI 686.01 (KIC 7906882) is a planetary candidate with orbital period 52.51 days and planet-to-star radius ratio $R_{\rm p}/R_\star=0.12$, orbiting around a G-type star with a {\em Kepler}-band magnitude of $13.58^{\rm m}$. The physical parameters of the host star are listed in Table~\ref{tab:k686-host}.

We assign epoch zero to the first transit observed by {\em Kepler}. KOI 686.01 has only one short cadence full transit at epoch +12. The out-of-transit variation amplitude of this transit is 851 ppm. After reducing the data according to Section~\ref{sec:data}, we fit this short cadence transit to a standard transit model, using the parameters given by {\em Kepler} as initials. With the best-fit parameters given by the standard transit model, we then set $f_\perp$ and $\theta_\perp$ free to allow the oblateness fitting. Our best-fit parameters for both standard and oblateness fittings, together with the initial parameters from {\em Kepler}, are listed in Table~\ref{tab:k686-fitting}. 

We show the posterior plots from the oblateness fitting in Figure~\ref{fig:k686-posterior}. Based on the best-fit parameters, the impact parameter $b_0$ of KOI 686.01 transiting the host is determined to be 0.64. For this $b_0$, the oblateness should be weakly constrained at $\theta_\perp=0$ and $\pm 90^\circ$ according to Figure~\ref{fig:b0-sina}. Therefore the result is consistent with our expectation.

No oblateness signature is detected for KOI 686.01. The overall projected oblateness $f_\perp$ can only be constrained to be $< 0.25$ ($68\%$ limit). For most $\theta_\perp \neq 0,\pm90^\circ$, we can rule out a planet more oblate than Saturn. For comparison, the posterior distributions of $f_\perp$ and $\theta_\perp$ for the simulated standard transit light curve are showed on the right panels of Figure~\ref{fig:k686-posterior}. The $68\%$ confidence upper limit of the projected oblateness for the injected null case is $0.26$. The comparison between the null case result and data-fit posterior suggests that KOI 686.01 is consistent with a spherical shape. However, since the upper limit we can set on the oblateness is very weak, an oblateness as large as that of Saturn can only be ruled out in $1-\sigma$ level if the planet is moderately tilted.

\subsection{KOI 197.01}

\begin{deluxetable}{cc}
\tablecaption{Physical parameters about the host star KOI 197 (KIC 2987027).
\label{tab:k197-host}}
\tablehead{Parameters & Values}
\startdata
\input{k197-physical.dat}
\enddata
\tablecomments{$^a$ Adopted from \citet{Santerne:2012}.}
\end{deluxetable}

\begin{deluxetable*}{cccc}
\tablecaption{Fitting parameters of KOI 197.
\label{tab:k197-fitting}}
\tablehead{Parameters &  Kepler & Standard model & Oblate model}
\startdata
\input{k197-fittings.dat}
\enddata
\end{deluxetable*}

\begin{figure*}
\centering
\plottwo{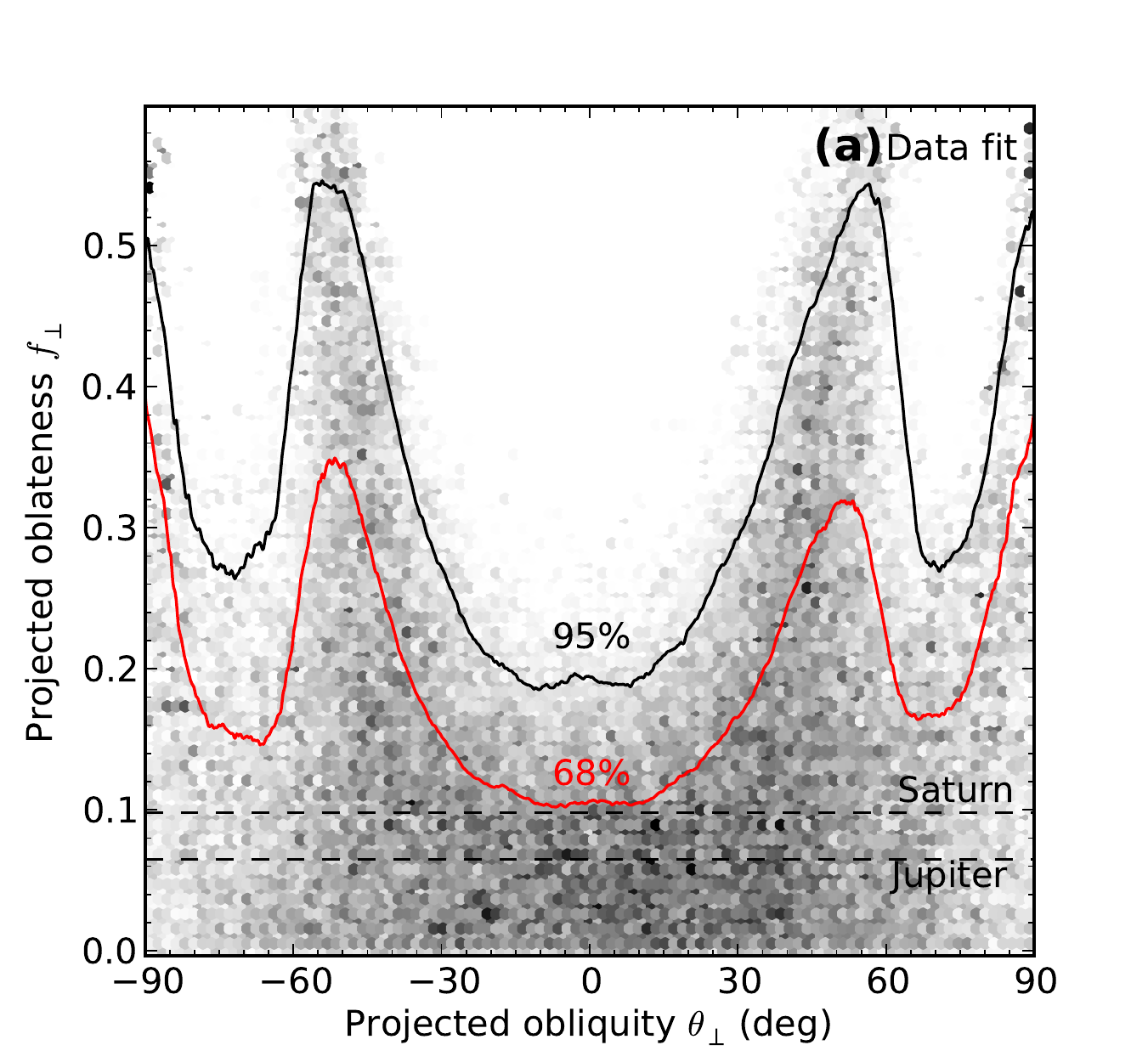}{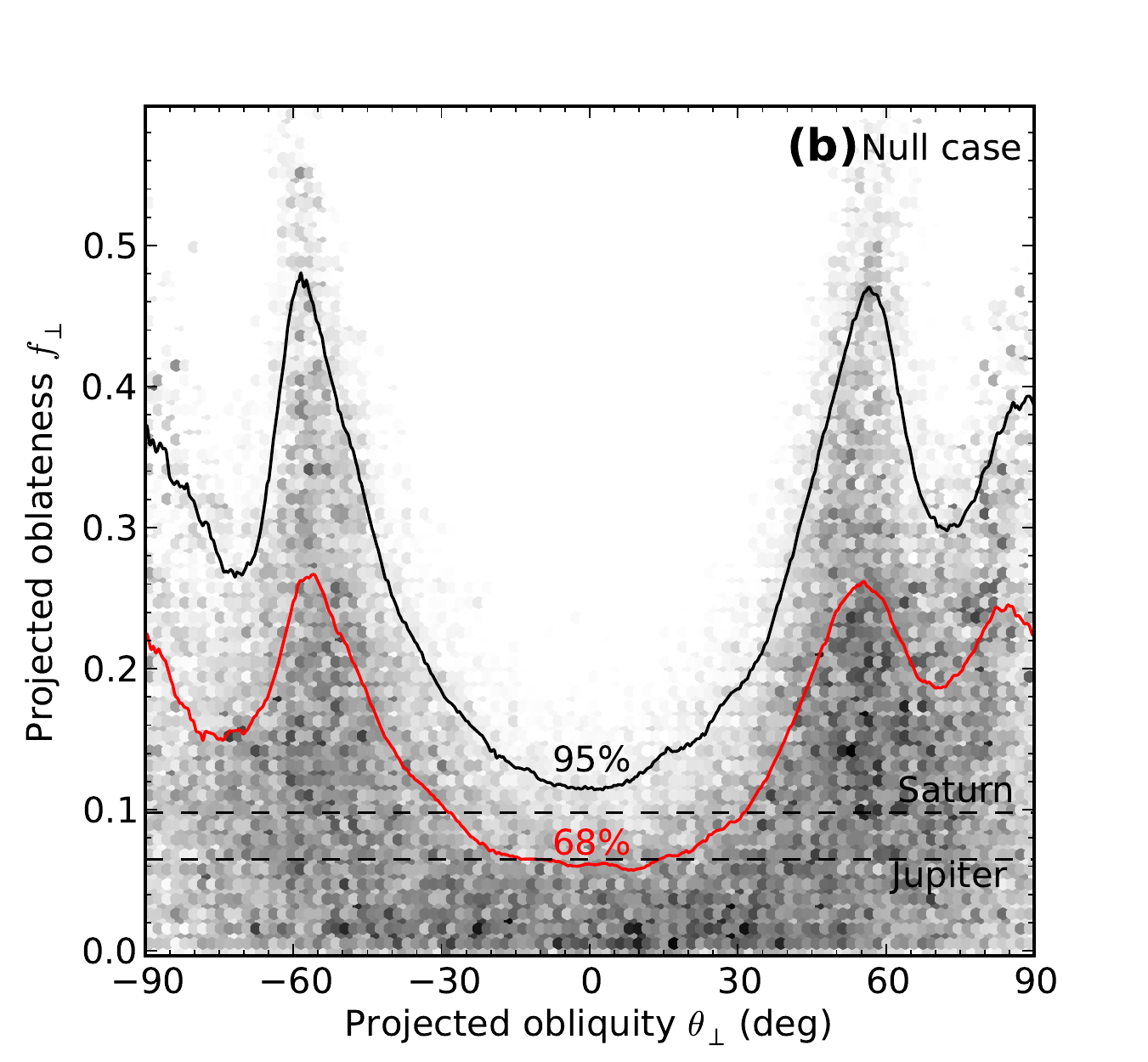}
\plottwo{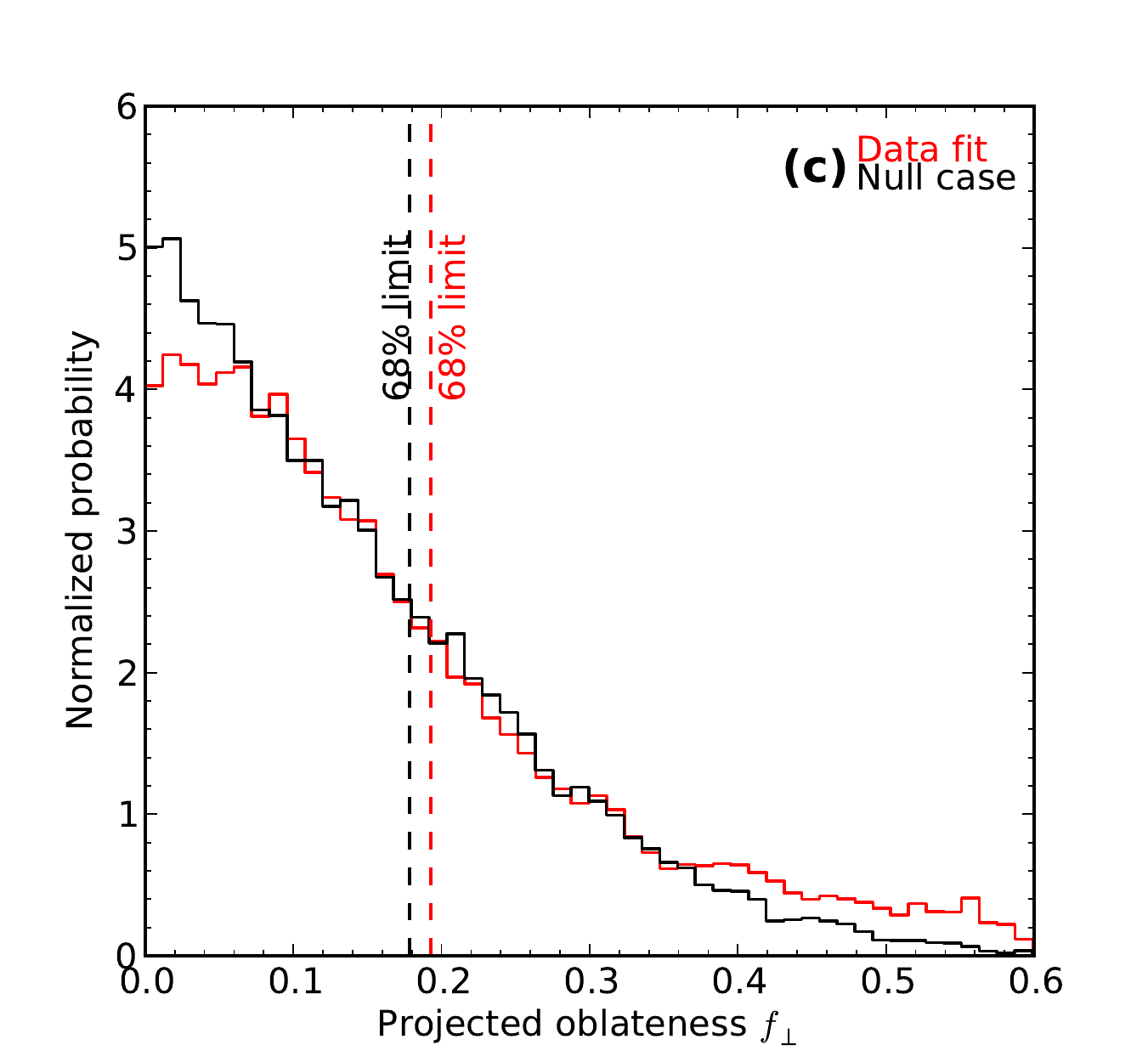}{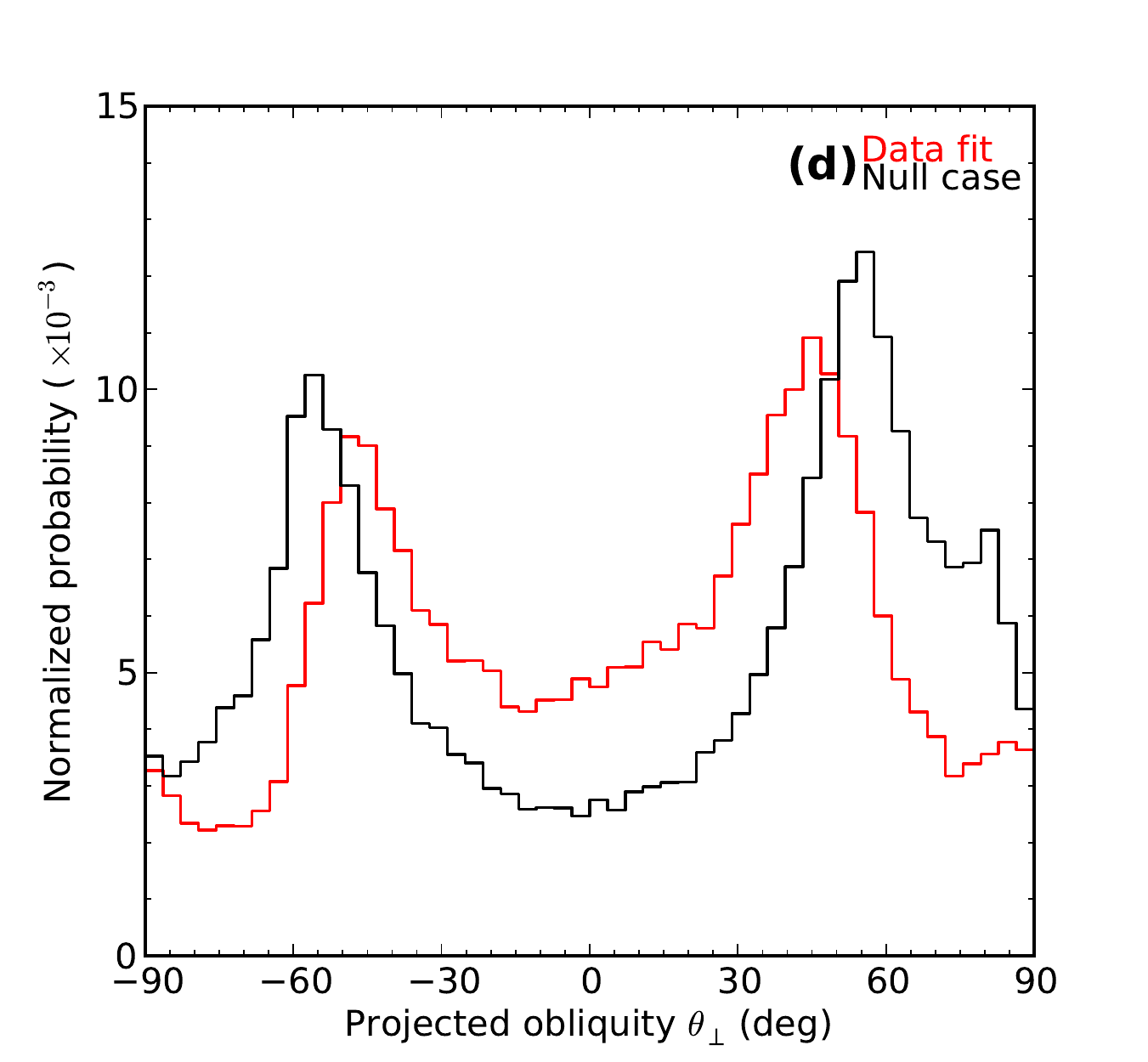}
\caption{The posterior distributions derived from the MCMC fitting of KOI-197.01. See caption of Figure~\ref{fig:k686-posterior} for details. 
\label{fig:k197-posterior}}
\end{figure*}

KOI 197.01 is a planetary candidate
\footnote{We notice that it was reported as a false positive based on secondary eclipse depth measured with Q0-Q2 data by \citet{Demory:2011}, but with data from Q1 -- Q16, we did not find the secondary eclipse at the phase reported by \citet{Demory:2011}.}
with orbital period 17.28 days and planet-to-star radius ratio $R_{\rm p}/R_\star=0.091$, orbiting around a K-type star whose {\em Kepler} band magnitude is $14.02^{\rm m}$. Physical parameters about the host star are listed in Table~\ref{tab:k197-host}.

There are 18 short cadence full transits of KOI 197.01 at epochs from +13 to +30. The averaged out-of-transit variation amplitude is 1196 ppm. After reducing the data according to Section \S\ref{sec:data}, we fit the standard transit model to the normalized light curves using the parameters given by {\em Kepler} as initials. With the best-fit parameters given by this standard transit model, we then set $f_\perp$ and $\theta_\perp$ free to allow the oblate model fitting. Our best-fit parameters for both standard and oblateness fittings, together with the initial parameters from {\em Kepler}, are listed in Table~\ref{tab:k197-fitting}. 

We show the posterior plots of the oblateness fitting in Figure~\ref{fig:k197-posterior}. Based on the best-fit parameters, the impact parameter $b_0$ of KOI 197.01 is 0.076. For this $b_0$, the oblateness is weakly constrained at $\theta_\perp \approx \pm 45^\circ$ according to Figure~\ref{fig:b0-sina}. Therefore the result is consistent with our expectation.

We detected no oblateness signal for KOI 197.01. The overall projected oblateness $f_\perp$ is constrained to be $< 0.19$ ($68\%$ limit). An oblateness as large as that of Saturn can be ruled out in $1-\sigma$ level if the projected obliquity $\theta_\perp$ is close to $0^\circ$. For comparison, the posterior distribution of $f_\perp$ and $\theta_\perp$ for the injected null-case standard transit light curves are shown on the right panel of Figure~\ref{fig:k197-posterior}. The $68\%$ confidence upper limit of the injected null case projected oblateness is $0.18$. We conclude that KOI 197.01 is consistent with a spherical shape, but with the oblateness only weakly constrained.

\subsection{KOI 423.01 (Kepler 39b)}

\begin{deluxetable}{cc}
\tablecaption{Physical parameters about the host star KOI 423 (KIC 9478990).
\label{tab:k423-host}}
\tablehead{Parameters & Values}
\startdata
\input{k423-physical.dat}
\enddata
\tablecomments{$^a$ Adopted from \citet{Bouchy:2011}.}
\end{deluxetable}

\begin{deluxetable*}{cccc}
\tablecaption{Fitting parameters of KOI 423.
\label{tab:k423-fitting}}
\tablehead{Parameters & Kepler & Standard model & Oblate model}
\startdata
\input{k423-fittings.dat}
\enddata
\end{deluxetable*}

\begin{figure*}
\centering
\epsscale{1.0}
\plottwo{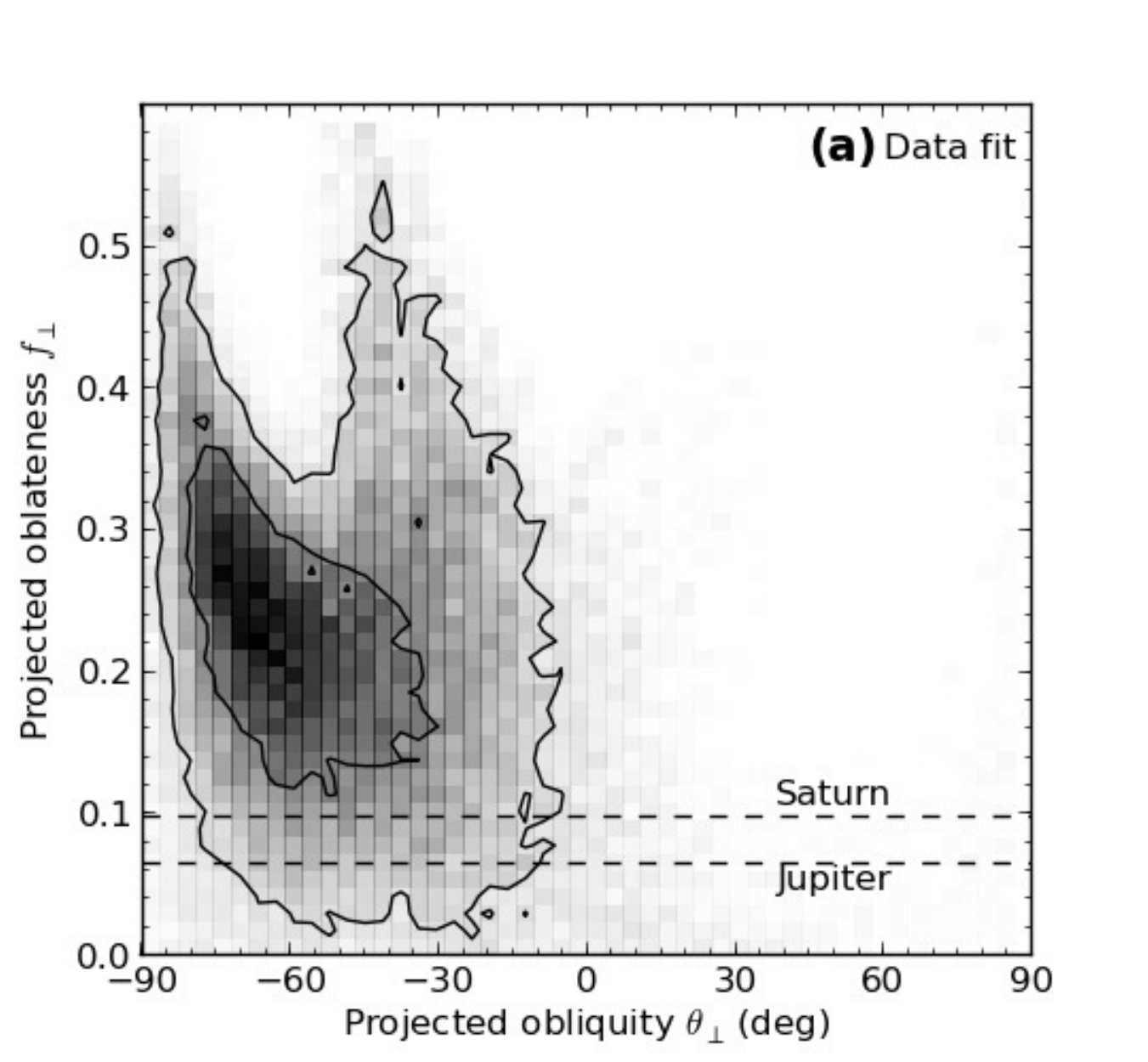}{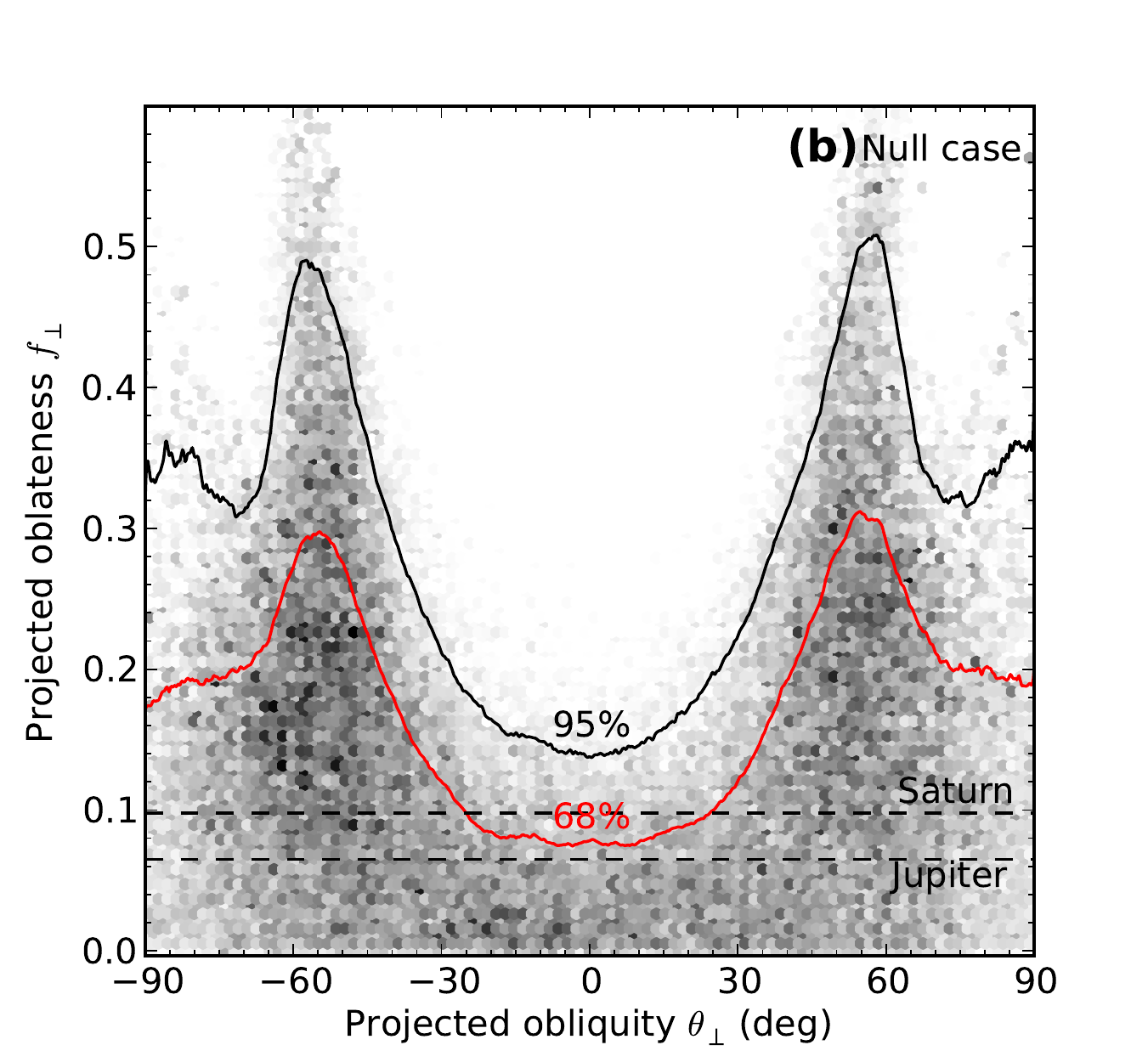}
\plottwo{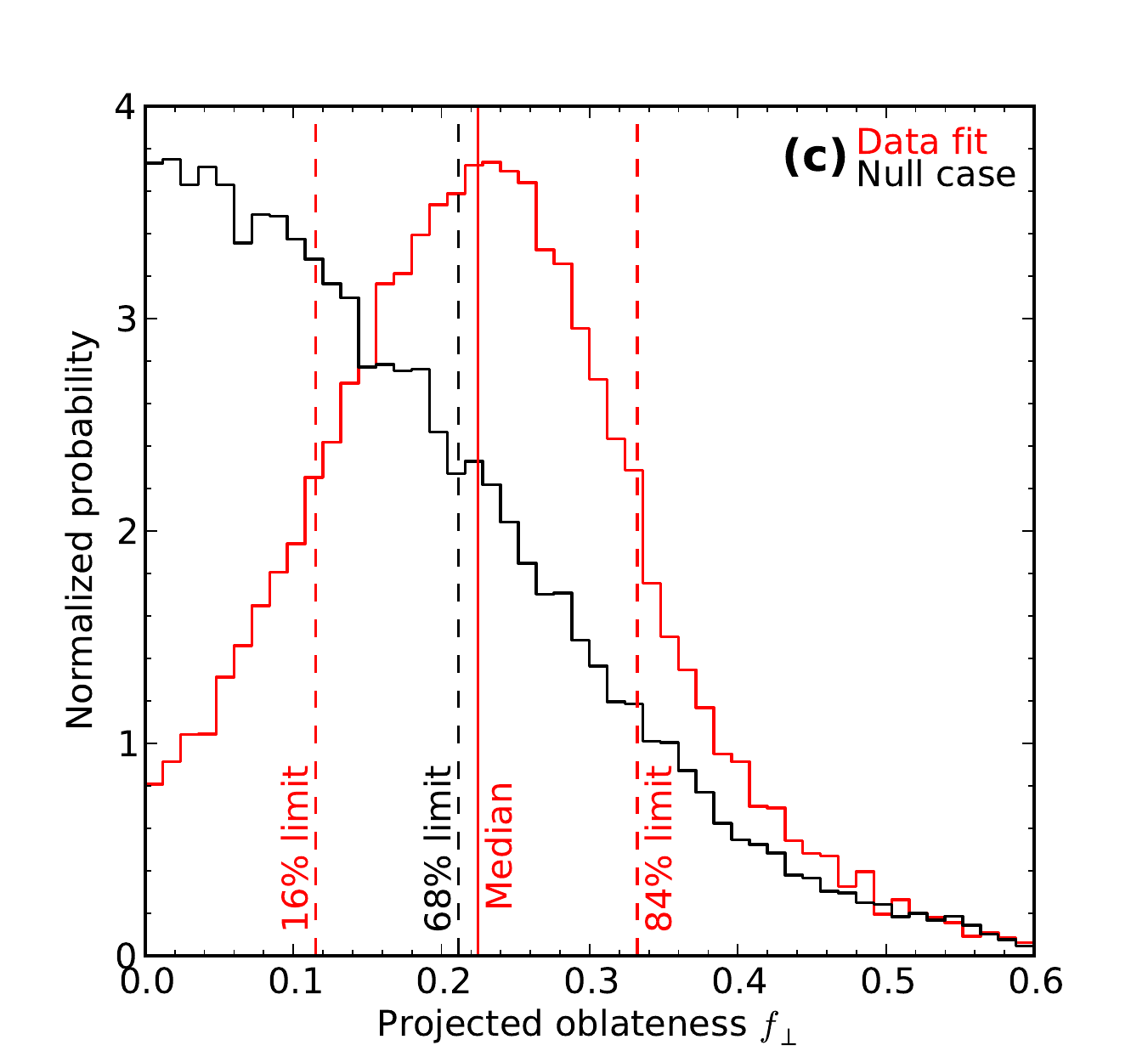}{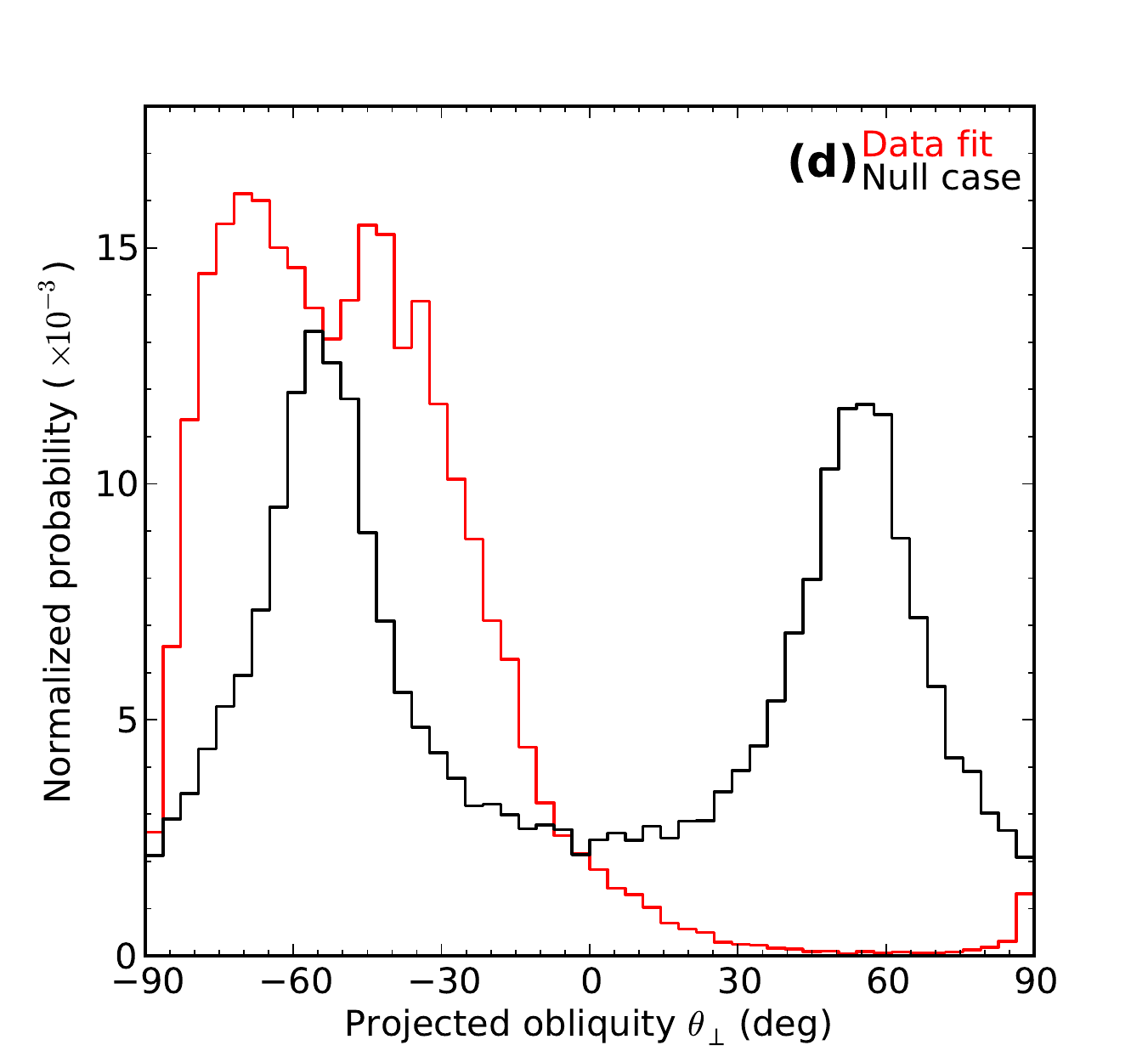}
\caption{The posterior distributions derived from the MCMC fitting of KOI-423.01. See caption of Figure~\ref{fig:k686-posterior} for details. The posteriors indicate a marginal detection of oblateness, with $f_\perp = 0.22_{-0.11}^{+0.11}$.
\label{fig:k423-posterior}}
\end{figure*}

\begin{figure*}
\centering
\begin{tabular}{cc}
  a) &\includegraphics[width=8cm]{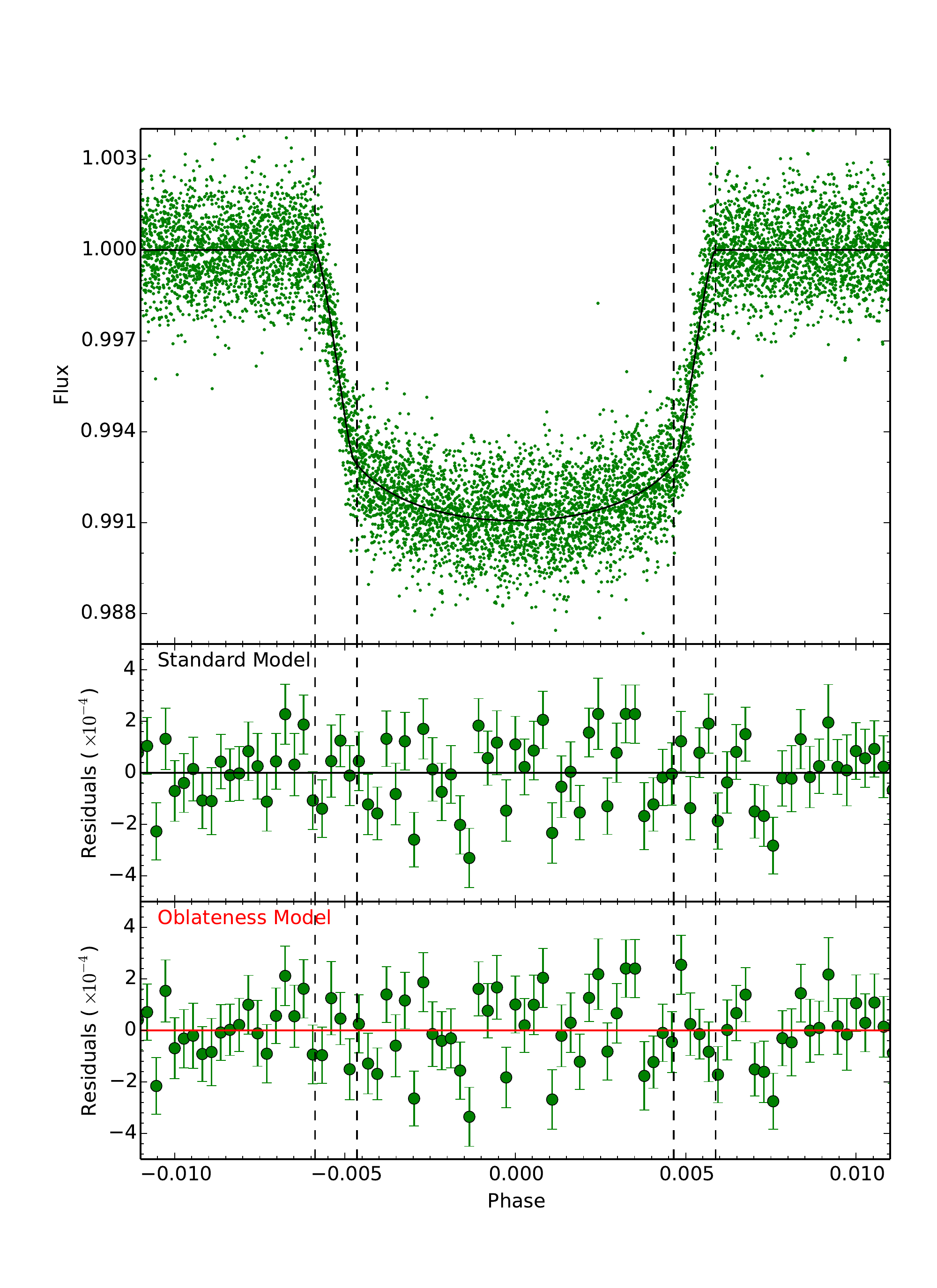}\\
  b) &\includegraphics[width=13cm]{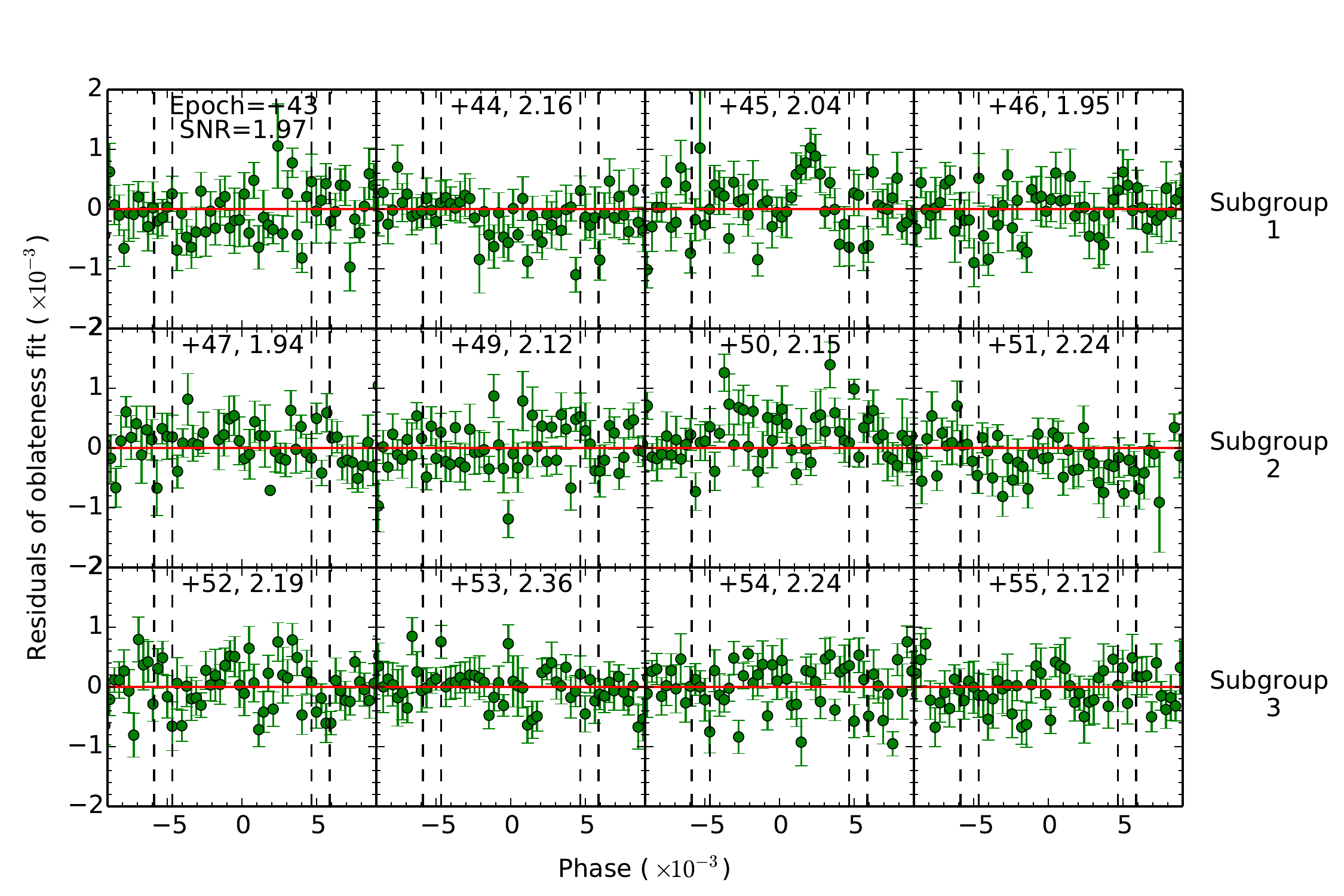}
\end{tabular}
\caption{a) The transits of KOI 423.01 are plotted in green (top panel). Residuals to the standard transit model (middle panel) and oblate transit model (bottom panel) are plotted. We only show the binned residuals and associated scattering for every 100 measurements. b) We plot the residuals to the oblate planet model of each individual transit. Ingresses and egresses are marked out by dashed lines. The transit epoches and oblate signal SNR (according to Equation~\ref{eq:snr}) are labeled. We only show the binned residuals and associated scattering for every 10 measurements.
\label{fig:k423-residuals}}
\end{figure*}

\begin{figure*}
\centering
\plottwo{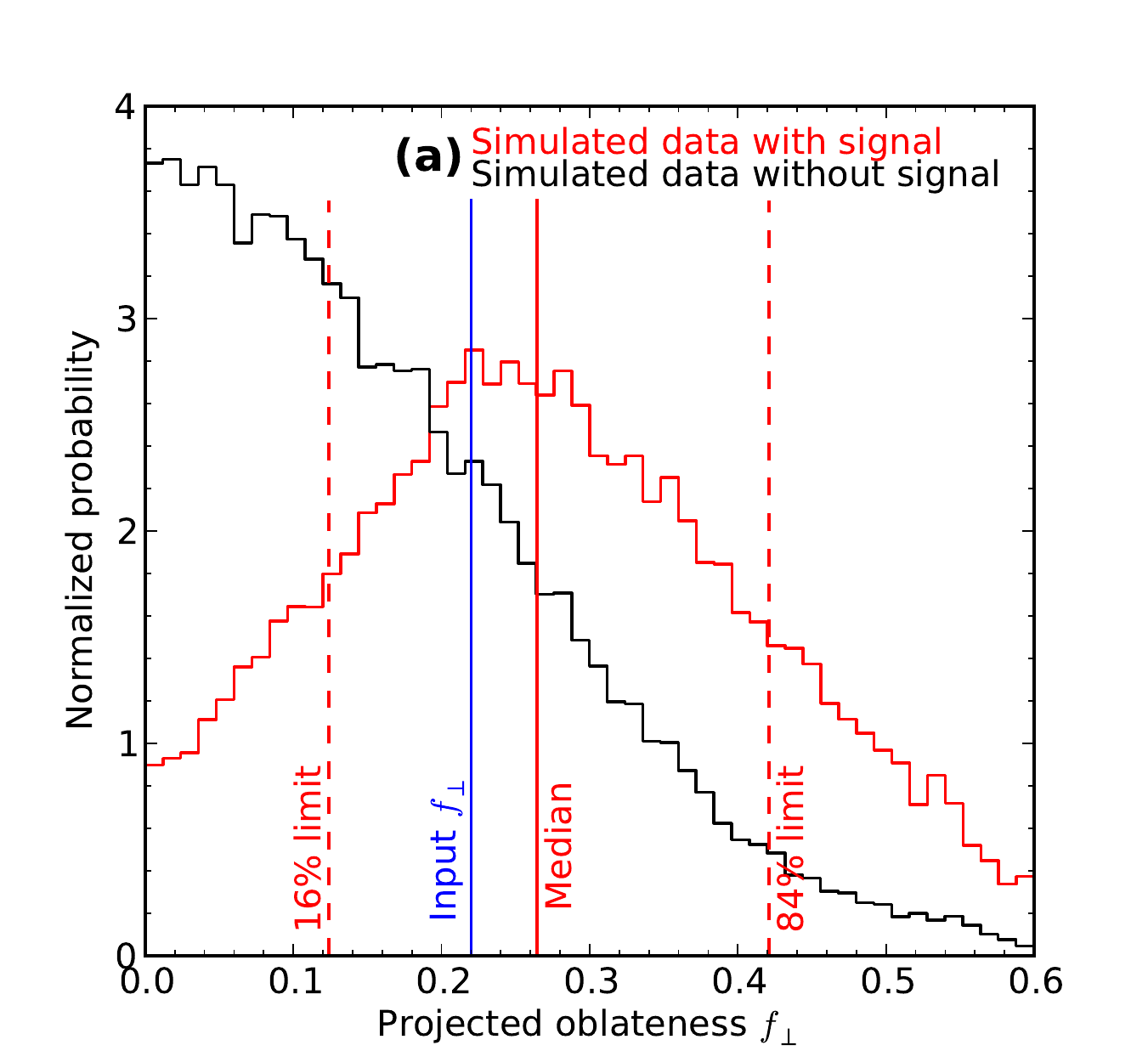}{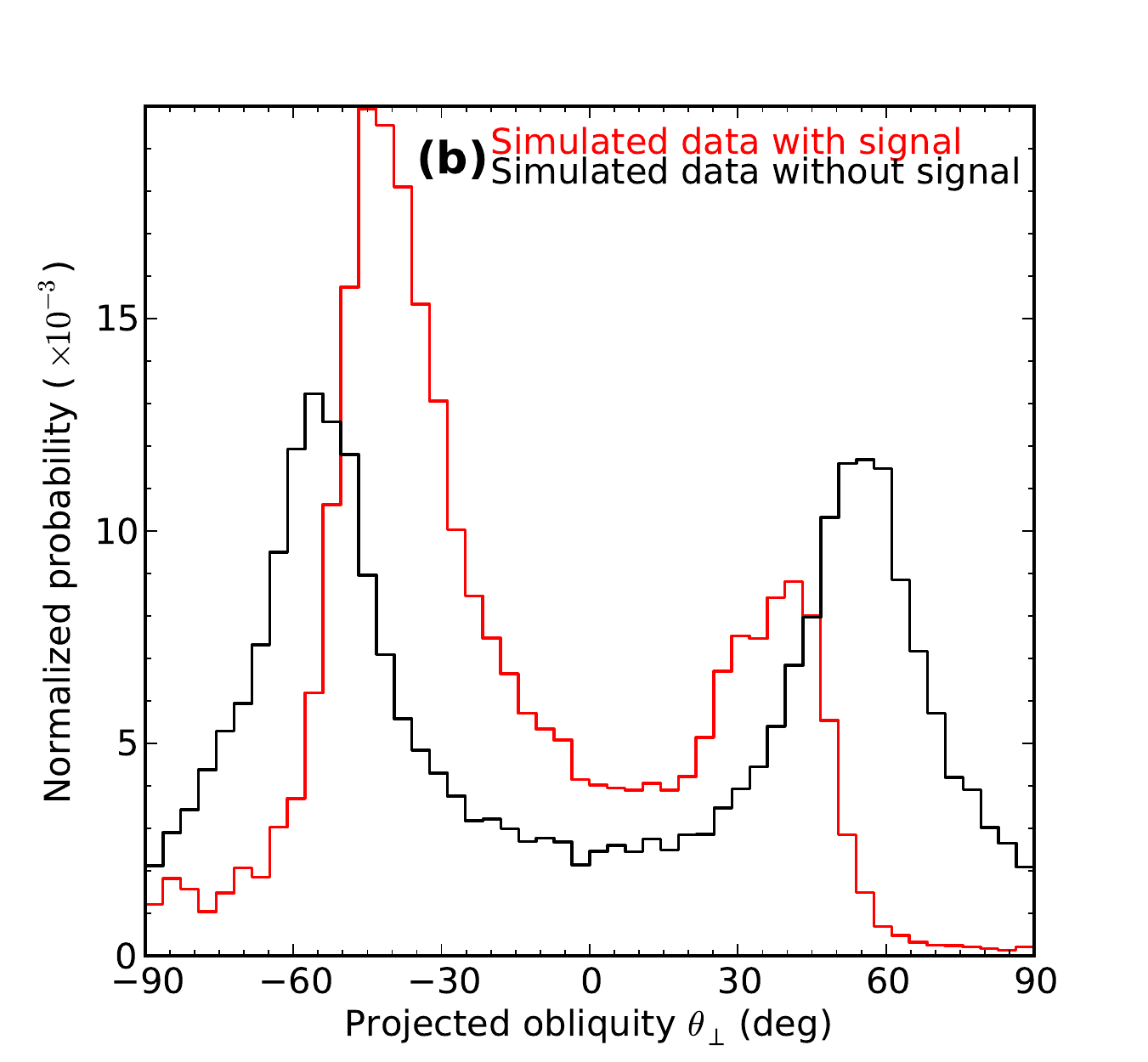}
\caption{Marginalized distributions of projected oblateness $f_\perp$ (\textit{left panel}) and obliquity $\theta_\perp$ (\textit{right panel}) for the simulated data with the $f_\perp=0.22$, $\theta_\perp=-40^\circ$ detected signal into the out-of-transit portions of the light curve (red), and for a spherical planet injected signal (black). 
\label{fig:k423-posterior-fk}}
\end{figure*}

\begin{figure*}
\centering
\plotone{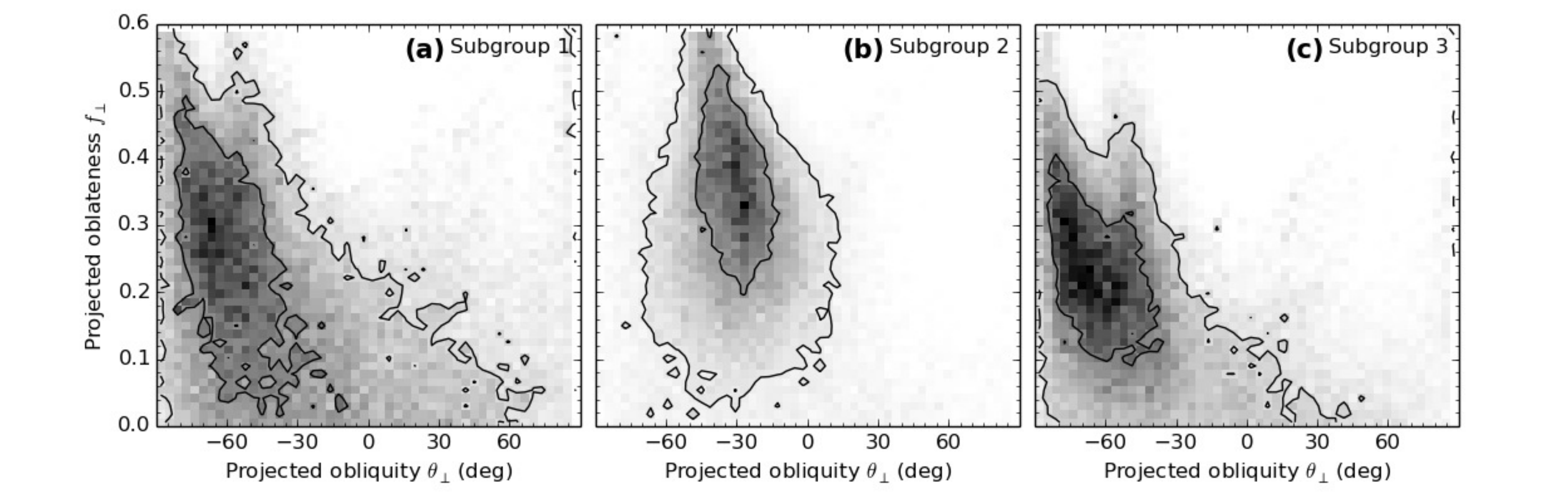}
\plotone{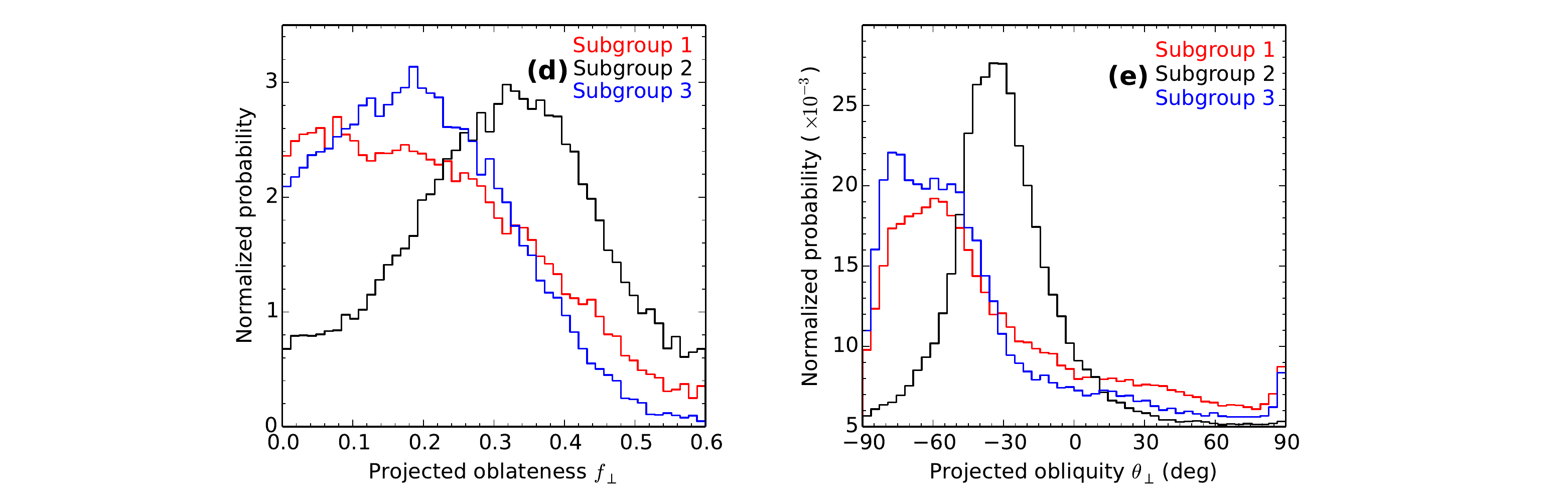}
\caption{Results of the subgroups fittings of KOI 423.01. 11 good short cadence transits, with epoch +45 excluded due to spot crossing event, are divided into three subgroups with three transits in the first subgroup and four in each of the other two. All transits are fitted simultaneously, sharing the same system parameters, except for $f_\perp$ and $\theta_\perp$.
\label{fig:423-subgroups}}
\end{figure*}

\begin{figure*}
\centering
\epsscale{0.9}
\plotone{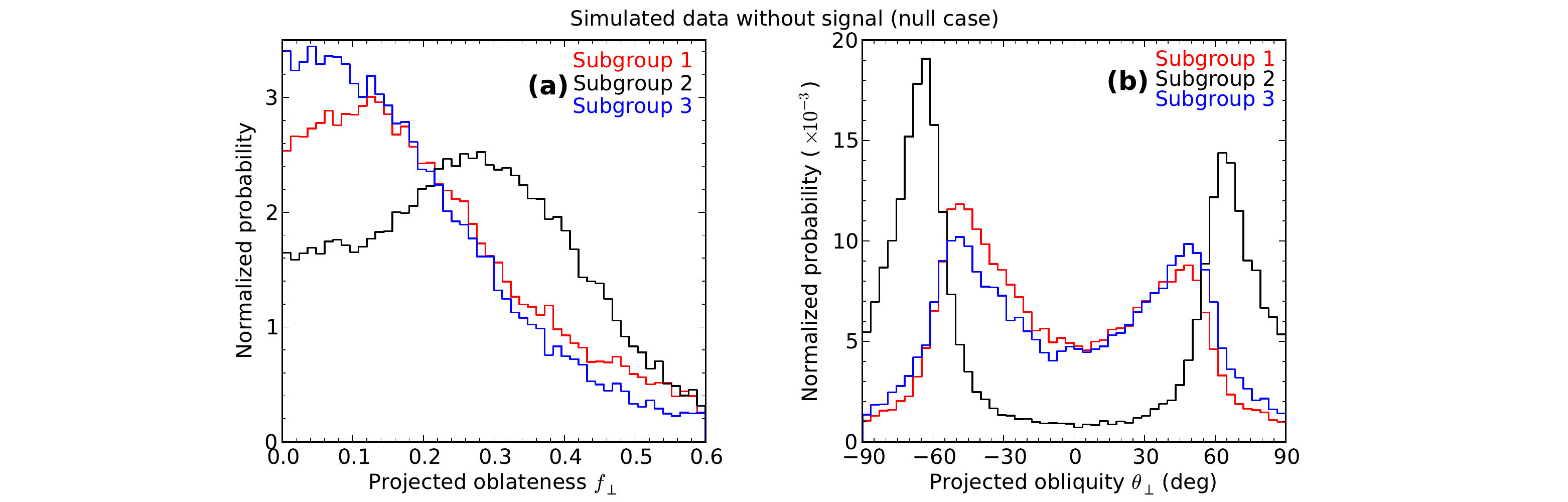}
\plotone{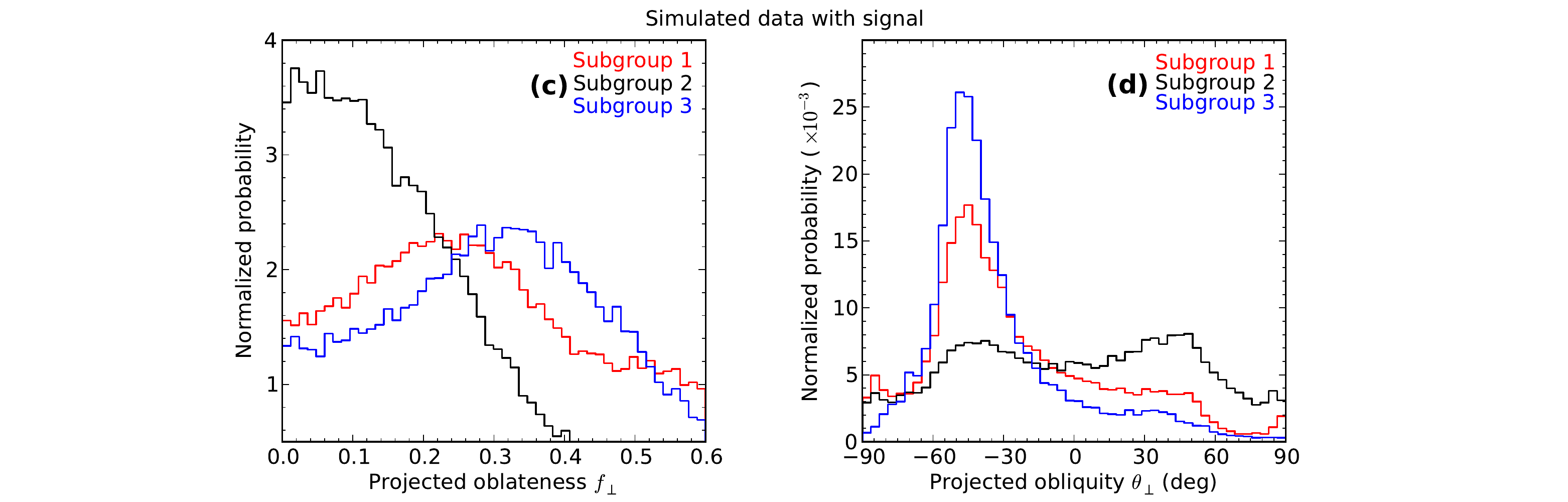}
\caption{Marginalized distributions of $f_\perp$ and $\theta_\perp$ for the subgroups fitting of simulated KOI 423.01 data with (\textit{Lower panel}) and without (\textit{Upper panel}) oblateness signal.
\label{fig:423-subgroups-fk}}
\end{figure*}

KOI 423.01 is a confirmed planet/brown dwarf (for simplicity, we just note it as a planet hereafter) with mass $18M_{\rm J}$ \citep{Bouchy:2011}. The planet has an orbital period of 21.09 days, orbiting around a F-type star which has a {\em Kepler}-band magnitude $14.33^{\rm m}$. The planet-to-star radius ratio of this system is $R_{\rm p}/R_\star=0.087$. Physical parameters about the host star are listed in Table~\ref{tab:k423-host}.

We use the 12 full short cadence transits of KOI 423.01 available, from epochs +43 to +55, with +48 absent, in our analysis. The averaged out-of-transit-variation amplitude is 1256 ppm. After reducing the data according to Section \S\ref{sec:data}, we fit the standard transit model to the normalized light curves using the parameters given by {\em Kepler} as initials. With the best-fit parameters given by this standard transit model, we then set $f_\perp$ and $\theta_\perp$ free to fit for oblateness. The best-fit parameters for both the standard and oblateness fittings, together with the initial parameters from {\em Kepler}, are listed in Table~\ref{tab:k423-fitting}.

The posterior plots from the oblateness fitting are shown in Figure~\ref{fig:k423-posterior}, the light curve and residuals of standard fit and oblateness fit are shown on the upper panel of Figure~\ref{fig:k423-residuals}. The residuals to each individual transit are shown in the lower panel of Figure~\ref{fig:k423-residuals}.

The marginalized distribution of $f_\perp$, shown in the lower left panel of Figure~\ref{fig:k423-posterior}, suggests that KOI 423.01 may have an oblateness of $f_\perp=0.22^{+0.11}_{-0.11}$, substantially larger than any planet in the Solar System. The total SNR of all 12 transits is 7.4 (Equation~\ref{eq:snr}). The SNR of each transit is also shown in the residual plots of Figure~\ref{fig:k423-residuals}.
 
To check for the robustness of such a signal, we first perform the injection and recovery of the null signal case, with a spherical planet, into the out-of-transit light curves. The recovered posteriors of the null signal case are plotted on the right panels of Figure~\ref{fig:k423-posterior}. The null detection case produces no oblateness signal, with the $f_\perp$ constrained to be $<0.2$ at the $1\sigma$ level.

We then inject a simulated oblate planet transit, with the same transit parameters as KOI 423.01, and $f_\perp=0.22$ and $\theta_\perp = -40^\circ$, into the out-of-transit sections of the light curve. The injected transits are shifted to the true transit epochs and fitted similar to the observed transits. The recovered posterior probability distributions are plotted in Figure~\ref{fig:k423-posterior-fk}. We successfully recovered the injected oblateness factor $f_\perp$, arriving at a posterior distribution similar to what we see with the observed data. This suggests that the observed oblateness signal is unlikely to be due to the red noise and stellar oscillations in the light curves.

To check for the consistency of such a signal between epochs, we choose 11 out of the 12 available observed transits, with epoch +45 excluded due to spot crossing event (see the lower panel of Figure~\ref{fig:k423-residuals}), and divide them into three subgroups with three transits in the first subgroup and four in each of the other two, to see if such a ``detection'' appears in each subgroup. We fit these subgroups simultaneously, such that all subgroups share the same system parameters, but each one has its own independent $f_\perp$ and $\theta_\perp$ values. In total, we fit 13 free parameters. The result of this fitting is shown in Figure~\ref{fig:423-subgroups}. Subgroups 2 and 3 both indicate oblateness detections and they are consistent within $1$-$\sigma$ level. Subgroup 1 does not show noticeable detection according to the marginalized posterior distribution of $f_\perp$, but the asymmetry in the marginalized distribution of $\theta_\perp$ as well as the 2D posterior contour between $f_\perp$ and $\theta_\perp$ suggests a potential oblateness detection, which is also consistent with that from the other two subgroups within $1$-$\sigma$ level.

However, we note that similar exercises performed on the simulated datasets can show inconsistency between epochs. As an example, shown in Figure~\ref{fig:423-subgroups-fk}, in the null case, one of the three subgroups indicates that a large oblateness is detected, and in the signal-injected case, one subgroup does not show any oblateness detection. These results suggests that the noise level per subgroup of transits is too high to allow for a robust consistency analysis. Therefore, we conclude that whilst we cannot validate the consistency of the observed oblateness signal based on the current data, we cannot dismiss such a potential signal either.

\section{Discussion}

In this study, we present the first search for rotationally oblate gas giants in the {\em Kepler} planet sample. The transit light curve of an oblate planet deviates from that of a spheroid primarily over the ingress and egress regions. We derived an analytical estimate for the amplitude of the deviation from the transit of a spheroid with the same transit parameters (Equation~\ref{eq:amplitude}). Through a signal injection and recovery exercise with the KOI-368.01 system, we showed that the signal of a Jupiter-sized planet with a Saturn-like oblateness can be detected with {\em Kepler} photometry at a high significance after only a single short cadence transit. 

We examined four selected \emph{Kepler} planets (candidates) for signatures of oblateness in their transit light curves. We find the hot-Jupiter HAT-P-7b to be consistent to a spheroid, with an overall oblateness constrained to be $<0.067$ at the 1-$\sigma$ level. In addition, an oblateness as small as that of Neptune (0.017) can be ruled out at 2-$\sigma$ confidence level if the planet is moderately inclined, although at some particular projected obliquities ($\theta_\perp = 0^\circ$ and $\pm 90^\circ$), an oblateness up to that of Saturn (0.1) is still allowed within the 2-$\sigma$ level. 

The oblatenesses of KOI 686.01 and KOI 197.01 are only weakly constrained due to the high noise level, with overall constraints of $<$$0.251$ and $<$$0.186$ for the $1\sigma$ cases, respectively. However, for most spin obliquities for two candidates, we can rule out oblateness larger than that of Saturn at the 1-$\sigma$ level.

KOI 423.01 (Kepler 39b) shows a potentially large oblateness ($f_\perp=0.22^{+0.11}_{-0.11}$). However, we find that the oblateness signal is only mildly self-consistent over multiple epochs, suggesting that the detection may not be robust. Given the long orbital period of $21.09$ days, and its eccentricity of 0.12, such an oblateness, if real, is likely to be rotationally, rather than tidally, induced. Using Equation~\ref{eq:rotation}, and the assumptions on the moment of inertia with Equation 8 of \citet{Carter:2010a}, we estimate the rotation period of KOI 423.01 to be $1.6\pm0.4$ hrs, assuming no line-of-sight obliquity. Whilst the derived rotation period is high in comparison with solar system gas giants, it is lower than the breakup rotation period of $\sim 0.9$ hrs for KOI 423.01. In addition, $\beta$ Pic b was recently measured to have a rotation rate of $25\pm3\,\text{km}\,\text{s}^{-1}$ \citep{Snellen:2014}. After cooling and contraction, the rotational period of the planet is expected to be $\sim 3$ hrs, similar to our predicted rotation period of KOI 423.01. We also note that KOI 423.01 has a larger-than expected radius \citep{Bouchy:2011,Mordasini:2012}. Its radius is larger by $20\%$ than both theoretical predictions and the radii of similar mass brown dwarfs \citep[e.g., CoRoT-3b,][]{Deleuil:2008}. Such a discrepancy cannot be solved by invoking an ad hoc increase in the atmospheric opacities because of the fact that KOI 423 is metal-poor \citep{Bouchy:2011}, or by tidal heating which only works in very eccentric orbits \citep{Dong:2013}. With the fact that a fast rotator is expected to be moderately inflated, we suggest that the discrepancy on KOI 423.01 could be explained by the oblateness measurement in this work.

Various other effects have the potential of systematically distorting the planet oblateness detection.  The gravity darkening effect can distort the transit light curve slightly, but the distortion appears mostly during the in-transit part of the light curve, and is only detectable for rapidly rotating stars \citep{Barnes:2011,Philippov:2013,ZhouHuang:2013}. Targets selected in this paper are too slowly rotating for gravity darkening to be an important factor. As for the influence of spot crossings, we note that even when the spot crossing occurs, it is much more likely to happen in the in-transit part than during the ingress/egress session, meaning that the detection of planetary oblateness is hardly affected. Furthermore, such events are in principle visually distinguishable and only affect a fraction of the transits, and thus can be excluded when one wants to do a finer analysis, as we have done in the KOI 423 case. Nonetheless, due to the high noise level of the light curve, we could not validate the oblateness detection of KOI 423.01\ in a high significance level, although the observed transits seem to show consistency between different epochs.

We interpret the detected oblateness of KOI 423.01 (Kepler 39b), if true, as being due to the preservation of its primordial spin angular momentum.  Although we cannot constrain the projected spin obliquity of KOI-423.01 very well, our results suggest that the spin axis of KOI-423.01 is significantly offset from the orbit normal. Whilst this is unexpected if the object is formed by accretion of gases from the protoplanetary disk, we recall that the gas giants in our Solar system are tilted to some extent, with the exception of Jupiter. To estimate the spin synchronization timescale of KOI 423.01, we adopt a planet mass of $18M_{\rm J}$ and an eccentricity of 0.12 \citep{Bouchy:2011}. The expected spin synchronization timescale for this brown dwarf is calculated to be $\simeq 320 Q^\prime_s$ yr, where $Q_s^\prime$ is the modified $Q$ value for synchronization. In order for it not to be tidally spun down within a time scale comparable or longer than the age of its host star (a few Gyr), its $Q^\prime_s$ value must be $\geq 10^7$ (see Appendix~\ref{sec:appendix-2} for the details of this estimation). 

The spin synchronization timescales for the three remaining planetary candidate targets, HAT-P-7b, KOI 686.01 and KOI 197.01, are $0.02 Q^\prime_s$ yr, $10 Q^\prime_s$ yr and $3.2 Q^\prime_s$ yr, respectively. Assuming that the ages of their host stars are comparable to that of the Sun and that their $Q^\prime _s$ values are in the range of that ($\sim 10^5$) estimated for Jupiter and Saturn \citep{YoderPeale:1981,Peale:1999}, these numbers suggest that HAT-P-7b, KOI 686.01 and KOI 197.01 have been tidally spun down, which is consistent with the non-detection of oblateness of these objects.

\acknowledgements

We would like to thank the anonymous referee for useful comments which helped to improve the manuscript. We also thank Christian Clanton and Daniel Bayliss for comments on the manuscript after carefully reading it through, and Andy Gould, Gaspar Bakos, Marc Pinsonneault, Scott Gaudi, Josh Winn and David Kipping for insightful discussions. Work by WZ was supported by NSF grant AST 1103471. This paper used data collected by the Kepler mission, funding for which is provided by the NASA Science Mission Directorate.

\appendix
\section{Derivation of the maximum amplitude} \label{sec:appendix-1}

\begin{figure}
\centering
\epsscale{0.8}
\plotone{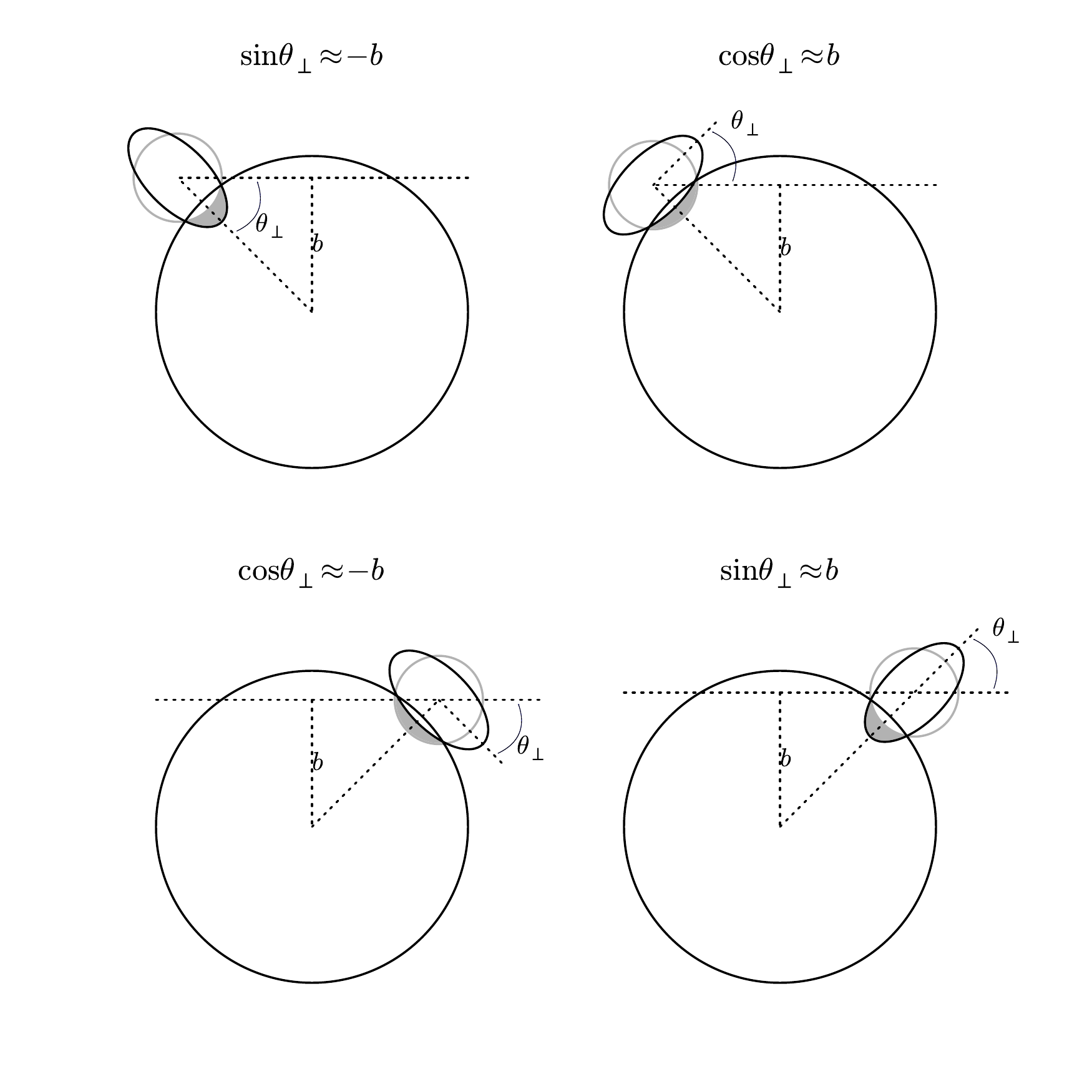}
\caption{Four cases where the maximum oblateness signal can be achieved for a specified impact parameter $b$. The shaded area is the maximum differential transit area between the oblate planet and a spherical planet with the same cross-sectional area.
\label{fig:max-signals}}
\end{figure}

The signal induced by an oblate planet with respect to the spherical planet with the same (projected) cross-section, reaches the maximum when the impact parameter $b$ and the projected spin obliquity $\theta_\perp$ have the relation depicted in any of the four cases in Figure~\ref{fig:max-signals}.

The induced signal arises from the differential transit area between the oblate planet and the spherical planet. In Figure~\ref{fig:max-signals} we mark the maximum differential area as shaded. From geometry we find this maximum differential area can be written as
\begin{equation}
\Delta S = r^2 \left( \arccos{\sqrt{\frac{1-f}{2-f}}} - \arctan{\sqrt{1-f}} \right) ,
\end{equation}
where $f$ is the measured oblateness of the planet, and $r=R_{\rm mean}/R_\star$ is the dimensionless mean radius. Therefore the differential transit signal, in the absence of limb darkening effect, is
\begin{equation}
{\rm Signal_{max}} = \frac{\Delta S}{\pi} = \frac{r^2}{\pi} \left( \arccos{\sqrt{\frac{1-f}{2-f}}} - \arctan{\sqrt{1-f}} \right).
\end{equation}
In the limit of $f \ll 1$, we find
\begin{equation}
{\rm Signal_{max}} = \frac{f r^2}{2\pi} \left[ 1+\frac{f}{2} + O(f^2) \right] .
\end{equation}

When the quadratic limb darkening effect is taken into account, the brightness profile is changed to
\[ I = 1-u_1(1-\sqrt{1-r^2})-u_2(1-\sqrt{1-r^2})^2 ,\]
where $u_1$ and $u_2$ are the quadratic limb darkening coefficients. Since this oblateness-induced signal happens at the ingress/egress part, the brightness of this differential area is therefore decreased by a factor of $(1-u_1-u_2)$. However, the total flux from the stellar plane also decreases from $\pi$ to
\begin{equation}
f = 2\pi \int_0^1 I rdr = \frac{\pi}{6} (6-2u_1-u_2) .
\end{equation}
Thus the amplitude of the oblateness signal decreases by a factor of $6(1-u_1-u_2)/(6-2u_1-u_2)$.

\section{Synchronization timescale estimation} \label{sec:appendix-2}
Dissipation of tidal perturbation by a star with a mass $M_\ast$ and radius $R_\ast$ on its planets leads to their orbital synchronization and circulation on time scales
\begin{equation}
\tau_\Omega = {Q_s ^\prime \alpha_p \over 9 \pi}
\left( {M_p \over M_\ast} \right) \left( {a \over R_p} \right)^3 
{\Omega_p / n  \over |f_3 (e) - f_4 (e) \Omega_p/n|} P_o,
\end{equation}
\begin{equation}
\tau_e = {Q_c^\prime \over 81 \pi}
{M_p \over M_\ast} \left( {a \over R_p} \right)^5 {18/11 \over
|f_2 (e) \Omega_p/n - 18 f_1(e)/11|} P_o
\end{equation}
where $\alpha_p$, $M_p$, $R_p$, $\Omega_p$, $n$, $a$, $e$, and $P_o$ are the planet's coefficient of moment of inertia, mass, radius, spin angular frequency, mean motion, orbital semi major axis, eccentricity and orbital period respectively. The magnitude of $f_1$, $f_2$, $f_3$, and $f_4$ reduces to unity for circular orbits, but they can be significantly larger for highly eccentric orbits\citep{DobbsDixon:2004}. The modified $Q^\prime = 3 Q/2 k$ where $k$ is the planet's Love number. In the equilibrium tidal model\citep{GoldreichSoter:1966}, the planets' Q-values for synchronization ($Q_s$) and circularization ($Q_c$) equal to each other. However, they may attain different values in dynamical tidal models\citep{OgilvieLin:2004}.

In the limit $\Omega_p > > n$,
\begin{equation}
{\tau_\Omega \over \tau_e} = {11 \alpha_p f_2 \over 2 f_4} {Q_s \over Q_c} 
\left( {R_p \over a} \right)^2 
{\Omega \over n}
\end{equation}
is much less than unity for short- and intermediate-period gas giants. This inequality leads to a possible state of pseudo synchronization in which
\begin{equation}
{\Omega_p \over n} = {f_3 \over f_4} =
{(1+15e^2/2+45e^4/8+15e^6/16) \over (1+3 e^2+3 e^4/8) (1 -e^2)^{3/2}}.
\end{equation}
It also implies that the circularization period (i.e., the period out to which the planets' orbits are circularized) is generally much longer than the synchronization time scale (i.e., the period out to which the planet's spins are synchronized with their orbital period).

Based on the value of $\alpha_p \simeq 0.2$ for Jupiter and Saturn \citep{Helled:2011a,Helled:2011b,HelledGuillot:2013} and the expression for $f_4$ \citep{DobbsDixon:2004}, we scale $\tau_\Omega  = \tau_9 Q_5 {\rm Gyr}$ where $Q_5 = Q_s^\prime/10^5$ and 
\begin{equation}
\tau_9 \equiv {10^{-4} \alpha_p \over 9 \pi} 
\left({M_p \over M_\ast}\right)
\left({a \over R_\ast}\right)^3 
{(1-e^2)^{9/2} \over (1+3e^2 + 3e^4/8)} {{P_o \over {\rm yr}}}.
\end{equation}

We adopt $M_p = 1.74 M_J$ for HAT-P-7b\citep{Winn:2009}, an upper limit of $1 M_J$ for KOI 686.01 based on the non-detection of radial velocity variations \citep{DawsonJohnson:2012, Diaz:2012}, $0.27 M_J$ for KOI 197.01 \citep{Santerne:2012} and  $18 M_J$ for KOI 423.01 \citep{Bouchy:2011}. We adopt $e=0$ for HAT-P-7b\citep{Pal:2008} and 0.62 for KOI 686.01 based on the analysis of \citet{DawsonJohnson:2012}. We also assume $e < < 1$ for KOI 197.01 and $e=0.12$ for 423.01.

With these model parameters, we obtain $\tau_9 =1.8 \times 10^{-6}$, $0.001$ and $3.2 \times 10^{-4}$ for HAT-P-7b, KOI 686.01 and KOI 197.01, respectively. These translate to spin synchronization timescales ($\tau_\Omega$) of 0.02 Myr, 1 Myr, and 0.32 Myr, respectively, assuming the $Q_s^\prime$ is of order $10^5$, which is in the range of that estimated for Jupiter and Saturn \citep{YoderPeale:1981, Peale:1999}. 

However, for the brown dwarf KOI 423.01, $\tau_\Omega = 160 Q^\prime _s$ yr.  The retention of its initial rapid spin requires $Q^\prime _s \geq 10^7$ which is two orders of magnitude larger than those for the planetary companion.  In a subsequent paper, we will explore the efficiency of dynamical tides induced by inertial waves as a possible cause for this dichotomy.

\end{CJK*}
\end{document}

%% file: k368-fitting.dat
$q_1$ & 0.27 & $0.26 \pm 1$ & $0.25 \pm 1$ & 0.27 & $0.25 \pm 1$ \\
$q_2$ & 0.22 & $0.23 \pm 5$ & $0.24 \pm 4$ & 0.22 & $0.25 \pm 5$\\
$P_{\rm orb}$ (days) & 110.321645 & $ 110.32167 \pm 2$ & $110.32166\pm 2$ & 110.321645 & $110.32166 \pm 2$\\
$t_0$ (BJD$_{\rm UTC}$-2454000) & 1030.36409  & $1030.3639 \pm 1$ & $1030.3639 \pm 1$ & 1030.36409 & $1030.3639 \pm 1$\\
$(R_{\rm p}+R_\star)/a$ & 0.02097 & $0.02103\pm 4$ & $0.02101\pm 5$ & 0.02097 & $0.02103 \pm 5$\\
$R_{\rm p}/R_\star$ & 0.08408 & $0.0842\pm 1$ & $0.0842 \pm 1$ & 0.08408 & $0.0842 \pm 2$\\
$i_0$ (deg) & 89.204 & $89.199\pm3$ & $89.200 \pm 4$ & 89.204 & $89.199\pm 4$\\
$f_\perp$ & 0.1 & -- & $0.11\pm3$ & 0 & $<0.03$\\

%% file: k2-physical.dat
$T_{\rm eff}$ (K) & $6350\pm80$ $^a$\\
$\log{g}$ (dex) & $4.07^{+0.04}_{-0.08}$ $^a$ \\
{\rm [Fe/H]} (dex) & $+0.26\pm0.08$ $^a$ \\
$M_\star\ (M_\odot)$ & $1.47^{+0.08}_{-0.05}$ $^a$ \\
$R_\star\ (R_\odot)$ & $1.84^{+0.23}_{-0.11}$ $^a$ \\
$v\sin{i_\star}\ ({\rm km/s})$ & $3.8\pm0.5$ $^a$ \\
$J$-band mag &9.555\\
$H$-band mag &9.344\\
$K$-band mag &9.334\\

%% file: k2-fittings.dat
$q_1$ $^{\rm a}$ & $0.28\pm2$ $^{\rm b}$ & $0.28\pm1$ & $0.28\pm1$ \\
$q_2$ $^{\rm a}$ & $0.33\pm1$ $^{\rm b}$ & $0.33\pm1$ & $0.33\pm1$ \\
$P_{\rm orb}$ (days) & $2.2047354 \pm 1$ & $2.2047350 \pm 3$ & $2.2047349 \pm 3$ \\
$t_0$ (BJD$_{\rm UTC}$-2454000) & $954.35780 \pm 2$ & $954.35860 \pm 2$ & $954.358608^{+16}_{-15}$ \\
$(R_{\rm p}+R_\star)/a$ & -- & $0.2595 \pm 4$ & $0.2597^{+16}_{-4}$ \\
$a/R_\star$ & $4.7$ & -- & -- \\
$R_{\rm p}/R_\star$ & $0.07545 \pm 2$ & $0.07748 \pm 4$ & $0.07750^{+9}_{-7}$ \\
$i_0$ (deg) & $88.2$ & $83.13 \pm 5$ & $83.11^{+9}_{-18}$ \\
$f_\perp$ & -- & -- & $<0.067$ \\

%% file: k686-physical.dat
$T_{\rm eff}$ (K) & $5559\pm71$ \\
$\log{g}$ (dex) & $4.47\pm0.3$ \\
{\rm [Fe/H]} (dex) & $-0.18$ \\
$M_\star\ (M_\odot)$ & $0.869\pm0.051$ \\
$R_\star\ (R_\odot)$ & $0.9\pm0.35$ \\
$J$-band mag &12.270\\
$H$-band mag &11.931\\
$K$-band mag &11.847\\

%% file: k686-fittings.dat
$q_1$ & $0.46\pm7$ & $0.29\pm15$ & $0.32^{+24}_{-11}$ \\
$q_2$ & $0.34\pm2$ & $0.39\pm29$ & $0.34^{+40}_{-26}$ \\
$t_0$ (BJD$_{\rm UTC}$-2454000) & $1004.6737 \pm 1$ & $1004.6743 \pm 2$ & $1004.6743 \pm 2$ \\
$(R_{\rm p}+R_\star)/a$ & -- & $0.0108 \pm 2$ & $0.0107^{+3}_{-2}$ \\
$a/R_\star$ & $109.3\pm4.2$ & -- & -- \\
$R_{\rm p}/R_\star$ & $0.1166 \pm 5$ & $0.121 \pm 2$ & $0.12 \pm 2$ \\
$i_0$ (deg) & $89.4$ & $89.60 \pm 2$ & $89.61^{+4}_{-2}$ \\
$f_\perp$ & -- & -- & $<0.251$ \\

%% file: k197-physical.dat
$T_{\rm eff}$ (K) & $4995 \pm 126$ $^a$\\
$\log{g}$ (dex) & $4.62 \pm 0.24$ $^a$ \\
{\rm [Fe/H]} (dex) & $-0.11 \pm 0.06$ $^a$ \\
$M_\star\ (M_\odot)$ & $0.77 \pm 0.09$ $^a$ \\
$R_\star\ (R_\odot)$ & $0.74 \pm 0.08$ $^a$ \\
$v\sin{i_\star}\ ({\rm km/s})$ & $11 \pm 1$ $^a$ \\
$J$-band mag &12.536\\
$H$-band mag &12.058\\
$K$-band mag &11.959\\

%% file: k197-fittings.dat
$q_1$ & $0.50\pm6$ & $0.41\pm6$ & $0.41^{+7}_{-6}$ \\
$q_2$ & $0.40\pm1$ & $0.48\pm7$ & $0.49^{+7}_{-6}$ \\
$P_{\rm orb}$ (days) & $17.276290 \pm 7$ & $17.27629 \pm 2$ & $17.27629 \pm 2$ \\
$t_0$ (BJD$_{\rm UTC}$-2454000) & $966.8386 \pm 2$ & $966.8395 \pm 4$ & $966.8396^{+3}_{-4}$ \\
$(R_{\rm p}+R_\star)/a$ & -- & $0.0301 \pm 4$ & $0.0299^{+7}_{-4}$ \\
$a/R_\star$ & $36.6\pm3.8$ & -- & -- \\
$R_{\rm p}/R_\star$ & $0.091 \pm 1$ & $0.0914 \pm 6$ & $0.0914^{+7}_{-4}$ \\
$i_0$ (deg) & $90.0$ & $89.88 \pm 28$ & $89.87^{+24}_{-25}$ \\
$f_\perp$ & -- & -- & $<0.186$ \\

%% file: k423-physical.dat
$T_{\rm eff}$ (K) & $6260\pm140$ $^a$\\
$\log{g}$ (dex) & $4.1\pm0.2$ $^a$ \\
{\rm [Fe/H]} (dex) & $-0.29\pm0.10$ $^a$ \\
$M_\star\ (M_\odot)$ & $1.10^{+0.07}_{-0.06}$ $^a$ \\
$R_\star\ (R_\odot)$ & $1.39^{+0.11}_{-0.10}$ $^a$ \\
$v\sin{i_\star}\ ({\rm km/s})$ & $16 \pm 2.5$ $^a$ \\
Age (Gyr) & $5.1 \pm 1.5$ $^a$ \\
$J$-band mag &13.236\\
$H$-band mag &12.999\\
$K$-band mag &12.918\\

%% file: k423-fittings.dat
$q_1$ & $0.40\pm4$ & $0.24\pm8$ & $0.23\pm5$ \\
$q_2$ & $0.27\pm1$ & $0.32\pm13$ & $0.34^{+11}_{-9}$ \\
$P_{\rm orb}$ (days) & $21.08717 \pm 2$ & $21.08722 \pm 4$ & $21.08722 \pm 3$ \\
$t_0$ (BJD$_{\rm UTC}$-2454000) & $1035.8578 \pm 3$ & $1035.8574 \pm 19$ & $0.8575 \pm 17$ \\
$(R_{\rm p}+R_\star)/a$ & -- & $0.0391 \pm 11$ & $0.0392^{+9}_{-11}$ \\
$a/R_\star$ & $29.6\pm3.5$ & -- & -- \\
$R_{\rm p}/R_\star$ & $0.086 \pm 1$ & $0.0888 \pm 8$ & $0.0889 \pm 6$ \\
$i_0$ (deg) & $90.0$ & $89.26 \pm 33$ & $89.25^{+30}_{-13}$ \\
$f_\perp$ & -- & -- & $0.22^{+11}_{-11}$ \\